\documentclass[12pt, letterpaper, oneside]{article}
\pdfoutput=1

\usepackage{graphicx}
\usepackage[body={6.5in, 9in}, right=1in, top=1in]{geometry}
\usepackage{amssymb}
\usepackage{amsmath} 
\usepackage{hyperref}
\usepackage[subnum]{cases}
\usepackage{color}

\newcommand\ea{\end{align}}
\newcommand\ba{\begin{align}}
\newcommand\ee{\end{equation}}
\newcommand\be{\begin{equation}}
\newcommand\eea{\end{eqnarray}}
\newcommand\bea{\begin{eqnarray}}
\newcommand{\sfrac}[2]{{\textstyle\frac{#1}{#2}}}
\newcommand\di{\partial}
\newcommand\pd{\partial}
\newcommand\mpl{M_{\rm Pl}}

\newcommand\ep{\epsilon}
\newcommand\dN{\stackrel{(1)}{\delta N}}
\newcommand\dNi{\stackrel{(1)}{\pd_i N^i}}

\numberwithin{equation}{section}

\begin{document}

\begin{center}
\LARGE{\textbf{Solid inflation}} \\[1cm]
\large{Solomon Endlich, Alberto Nicolis, 
and Junpu Wang}
\\[0.4cm]

\vspace{.2cm}
\small{\textit{Department of Physics and ISCAP, \\ 
Columbia University, New York, NY 10027, USA}}

\end{center}

\vspace{.2cm}

\begin{abstract}
We develop a cosmological model where primordial inflation is driven by a `solid', defined as a system of three derivatively coupled scalar fields obeying certain symmetries and spontaneously breaking a certain subgroup of these. The symmetry breaking  
pattern differs drastically from that of standard inflationary models: time translations are unbroken. This prevents our model from fitting into the standard effective field theory description of adiabatic perturbations, with crucial consequences for the dynamics of cosmological perturbations. Most notably, non-gaussianities in the curvature perturbations are unusually large, with $f_{\rm NL} \sim 1/(\epsilon c_s^2)$, and have  a novel shape: peaked in the squeezed limit, with anisotropic dependence on how the limit is approached.
Other unusual features include the absence of adiabatic fluctuation modes {\em during} inflation---which does not impair their presence and near scale-invariance after inflation---and a slightly {\em blue} tilt for the tensor modes.

\end{abstract}

%%%%%%%%%%%%%%%%%%%%%%%%%%%%%%%%%%
%%%%%%%%%%%%%%%%%%%%%%%%%%%%%%%%%%
\section{Introduction}

There is certainly no shortage of models for primordial inflation. We regret to inform the reader that---as anticipated in the Abstract---we are going to add our own to this list. However, we feel that the inflationary model we introduce here presents conceptually novel features that make it stand out as a radical alternative to the standard inflationary scenario. 
The main reason for this is that our model does  not conform to the standard symmetry breaking pattern of inflationary models, and this has far-reaching implications. 

In the usual cases, the matter fields $\psi_m$ feature time-dependent cosmological background solutions $\bar \psi_m(t)$, which spontaneously break time translations. As a result, there is one fluctuation mode $\pi(x)$ that can be identified with the associated Goldstone excitation. 
Roughly speaking, it can be thought of as an in-sync perturbation of all the matter fields, of the form
\be
\psi(x) =  \bar \psi_m(t + \pi(x)) \simeq  \bar \psi_m(t ) + \di_t{\bar \psi}_m(t) \cdot \pi(x) \; .
\ee
When coupling to gravity is taken into account, such a mode describes adiabatic perturbations. As usual for Goldstone bosons, the spontaneously broken symmetry puts completely general, non-trivial constraints on these perturbations' dynamics  \cite{CLNS}. This property is at the basis of the model-indepedent approach that goes under the name of ``effective field theory of inflation'' \cite{CCFKS}, whose tenets are particularly compelling since they encompass---at first glance---{\em all} cosmological models: cosmology {\em is} about time-dependent, homogeneous, and isotropic field configurations.

However, as we will see, there are other possibilities. In our case, we will be dealing with matter fields featuring {\em time-independent}, {\em $\vec x$-dependent} background solutions. Apparently, this contradicts two facts about inflationary cosmology:
\begin{enumerate}
\item
The universe is homogeneous and isotropic;
\item
In an expanding universe physical quantities depend on time and, more to the point, that one needs a physical `clock'---a time-dependent observable---to tell the universe when to stop inflating.
\end{enumerate}
As for item 1: $\vec x$-dependent solutions {\em can} be compatible with the homogeneity and isotropy we want for cosmological solutions and for the dynamics of their perturbations,  provided extra symmetries are imposed. For instance, to get an FRW solution for the gravitational field, we need an homogeneous and isotropic background stress-energy tensor. This can arise from
matter fields that are not homogeneous nor isotropic, provided there are internal symmetries acting on the fields that can reabsorb  the variations one gets by performing translations and rotations. The simplest example is that of a scalar field with a vacuum expectation value
\be \label{phi=x}
\langle \phi \rangle = \alpha x \; .
\ee
Such a configuration breaks translations along $x$. However, if one postulates an internal shift symmetry $\phi \to \phi + a$, then the configuration above is invariant under a combined spacial translation/internal shift transformation. As we will see, this is enough to make the stress-energy tensor and the action for small perturbations invariant under translations. To recover isotropy as well, one needs more fields, and more symmetries. For instance---in fact, this is the case that we will consider in this paper---one can use three scalar fields $\phi^I(x)$ ($I=1,2,3$), with internal shift and rotational symmetries
\begin{align}
\phi^I & \to \phi^I + a^I \; , \qquad a^I = {\rm const} \; ,  \label{shift}\\
\phi^I & \to O^I {}_J \phi^J \; , \qquad O^I {}_J \in SO(3) \label{rot} \; ,  
\end{align}
so that the background configurations
\be \label{vev}
\langle \phi^I \rangle = \alpha x^I 
\ee
are invariant under combined spacial translation/internal shift transformations, and under combined spacial/internal rotations. As we will review in  sect.~\ref{solids}, such a system has the same dynamics as those of the mechanical deformations of a solid---the phonons. In this sense,  the  cosmological model that we are putting forward  corresponds to having a solid driving inflation. If this interpretation causes the reader discomfort---in particular, if having a solid that can be stretched by a factor of $\sim e^{60}$ without breaking sounds implausible---one should think of our model just as a certain scalar field theory. As we will see, the structure of such a theory is the most general one compatible with the postulated symmetries---and the impressive stretchability we need can also be motivated by an approximate symmetry---so that from an effective field theory standpoint, ours is  a perfectly sensible inflationary model. From this viewpoint, the fact that the solids we are used to in everyday life behave quite differently---quantitively, not qualitatively---seems to be an accident: they lack the `stretchability symmetry'.

As for apparent contradiction number 2: In our model the role of the physical clock will be played by the metric. More precisely, it will be played by (gauge invariant) observables, made up of our scalars  and of the metric, like for instance the energy density or  the pressure. These can depend on time even for purely space-dependent scalar backgrounds, because in the presence of a non-trivial stress-energy tensor, the metric will depend on time, in a standard FRW fashion. Doesn't this correspond to a spontaneous breakdown of time translations too? At some level it is a matter of definition, but we will argue in sect.~\ref{clock} that the operationally useful answer is `no', in the sense that there is no associated Goldstone boson, and  that one cannot apply to our case the standard construction of the effective field theory of inflation as given in \cite{CCFKS}.

\begin{figure}[t]
\begin{center}
\includegraphics[width=16cm]{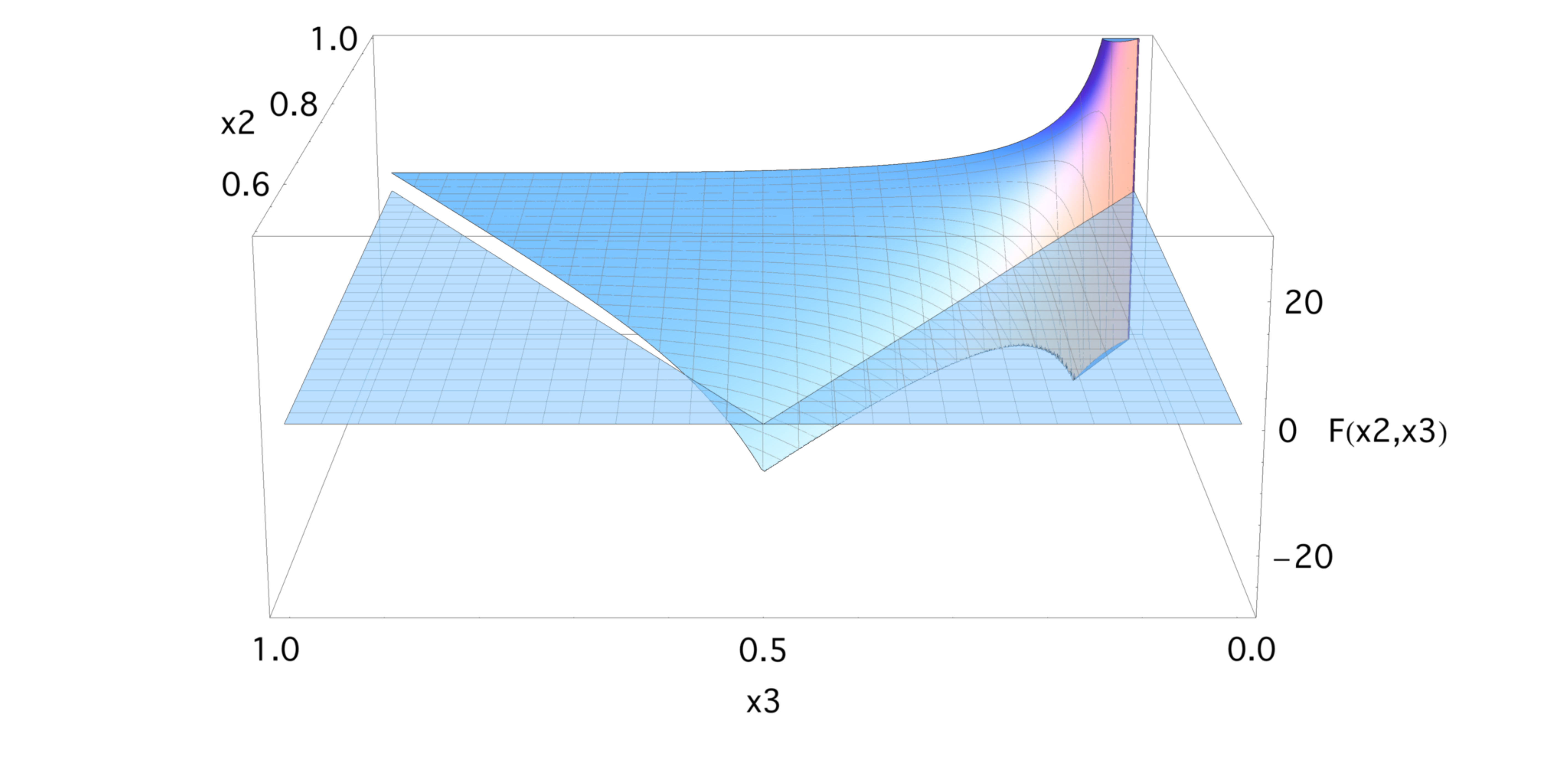}
\caption{\small \it The shape of non-gaussianities for our model, according to the standard conventions and definitions of ref.~\cite{BCZ}. \label{shape}}
\end{center}
\end{figure}

Formal considerations aside, our peculiar symmetry-breaking pattern has concrete physical implications, with striking observational consequences. For instance, it predicts a three-point function for adiabatic perturbations with a `shape' that is not encountered in any other model we are aware of.  Without going into details here, we display it in fig.~\ref{shape} for the benefit of the non-gaussianity aficionados: it diverges in the squeezed limit, but in a way that depends on the {\em direction} in which one approaches the limit, with a quadrupolar angular dependence. Its overall amplitude is also unusually large, corresponding to $f_{\rm NL} \sim \frac{1}{\epsilon} \frac{1}{c^2_s}$.

Before proceeding to spelling out our model in detail, we close this Introduction with two qualifications.
The first is that we have been using (and will be using)  a somewhat misleading, but fairly standard, language: when  spontaneously broken symmetries are gauged, the associated would-be Goldsone bosons are not in the physical spectrum---rather, they are `eaten' by the longitudinal polarizations of the gauge fields. In fact, there is a gauge, the so-called unitary gauge, in which the Goldstone fields are set to  zero. In our case we will be dealing with spontaneously broken translations and rotations, and when gravity is dynamical, these are gauged. In unitary gauge one can set the scalars to their vevs \eqref{vev}, and have the corresponding excitations show up in the metric. The Goldstone language is still useful though, in that it captures the correct high energy/short distance dynamics of these excitations. For massive gauge theories, this statement goes under the name of ``the equivalence theorem''. For cosmological models, it is just the statement that at sub-cosmological distances 
and time-scales, in first approximation one can neglect the mixing between matter perturbations and gravitational ones.
We hope the Goldstone boson nomenclature will be more useful than misleading.

The second qualification is that our model is not entirely new.
A different formulation of essentially the same inflationary model was put forward and briefly analyzed by Gruzinov in \cite{Gruzinov}. Our emphasis here will be on the peculiar symmetry breaking pattern, on the effective field theory viewpoint, on a systematic analysis of cosmological perturbations---including their non-gaussian features---and in general on the conceptual and technical differences with more standard inflationary models. Cosmological solids have also been used as an exotic model for dark matter \cite{BS}, and more recently as dynamical media with negative pressure but well-behaved excitations \cite{GHKRT}.

As a general guide for the reader, the bulk of our (long) paper can be schematically divided into three parts:
\begin{itemize}
\item
sects.~\ref{solids}--\ref{clock} introduce our model and discuss its field-theoretical and conceptual aspects, including why it cannot fit into the standard EFT of inflation; 
\item
sects.~\ref{perturb}--\ref{The size and shape of non-gaussianities} contain a technical analysis of cosmological perturbations, up to their three-point correlation function; 
\item
sects.~\ref{Why is zeta not conserved?} and \ref{reheating} address a number of conceptual subtleties in the analysis of cosmological perturbations, stemming from a very unusual feature of our model: the absence of {adiabatic} perturbations during inflation.
\end{itemize}
A brief summary of our results is contained in sect.~\ref{conclusions}, along with a number of possible generalizations of our scenario. Much technical material is contained in the Appendix.

%%%%%%%%%%%%%%%%%%%%%%%%%%%%%%%%%%
%%%%%%%%%%%%%%%%%%%%%%%%%%%%%%%%%%
\section{Effective field theory for solids (and fluids)}\label{solids}
To begin with, let us review how one can describe the mechanical degrees of freedom of a solid in modern effective field theory terms. The first systematic approach to this question has probably been that of ref.~\cite{Leutwyler}. Here instead we will review the equivalent construction of \cite{DGNR, ENRW} (see also \cite{Son}), whose notation we will follow. 

Consider a medium, filling space. Let's neglect for the moment gravitational effects and let's take  the metric to be flat. The medium's configuration space can be parameterized by its volume elements' individual positions. We can attach a (three-dimensional) comoving label $\phi^I$ (with $I=1,2,3$) to each volume element, and follow the volume element's trajectory in physical space, as a function of its comoving label and of time, $\vec x (\phi^I, t)$. We can also do the opposite, which for our purposes turns out to be  more convenient: since at any given time the mapping between $\phi^I$ and $\vec x$
is invertible, one can parameterize the system via $\phi^I(\vec x, t)$, that is, by keeping track of the comoving labels of the volume elements going through physical position $\vec x$, as a function of time. This is a completely equivalent description, yet it has the advantage of implementing the spacetime symmetries in the usual way: $(\vec x, t) = x^\mu$ are just the usual Minkowski coordinates, and the $\phi^I$'s are three Poincar\'e-scalars. Our problem then is reduced to constructing a relativistic low-energy effective field theory for three scalar fields  in Minkoswki space obeying certain internal symmetries, which we will discuss in a moment.
To avoid future confusion with standard cosmological nomenclature,  in the following we will call the $\phi^I$'s `internal' coordinates, and reserve `comoving' for the standard FRW coordinates for when we introduce gravity.

Now, apart from faithfully following the volume elements they are associated with, the internal coordinates are completely arbitrary. There is a coordinate system however that makes discussing internal symmetries---the symmetries that act on the $\phi^I$ fields---particularly easy: it is the choice that reduces to \eqref{vev} for a medium at rest, in equilibrium, in an homogeneous state at given external pressure. The parameter $\alpha$ measures the compression level, which can be dialed physically  by changing the external pressure. Then, since we are supposed to be describing an homogeneous medium, but the expectation values \eqref{vev} break translations, we are forced to postulate the existence of internal shift symmetries
\be \label{shift2}
\phi^I \to \phi^I + a^I  \; , \qquad a^I = {\rm const} \; ,
\ee
so that there is an unbroken linear combination of spacial and internal translations that mimics the physical homogeneity we are trying to reproduce. In other words, thanks to  the symmetry \eqref{shift2}, the internal space of the medium is always homogeneous. Whether this homogeneity is realized in physical $\vec x$-space depends on the medium's configuration, or state. For a state like \eqref{vev}, it is.
This logic is, of course, the same as we were employing in the Introduction to motivate our cosmological model.

We may want to impose rotational symmetries. The background configuration \eqref{vev} breaks spacial rotations. For a crystalline solid, like e.g.~one with a cubic lattice, a discrete subgroup of these should be preserved. This can be achieved by demanding an internal symmetry
\be
\phi^I \to O^I {}_J \phi^J
\ee
for a generic matrix $O$ belonging to the desired subgroup of $SO(3)$, so that the background \eqref{vev} is invariant under combined spacial and internal rotations. 
One can also impose full $SO(3)$ invariance, in which case one would be describing an isotropic solid with no preferred axes---a `jelly'.
For simplicity, in this paper we will consider this $SO(3)$-invariant case only. We will come back to comment briefly on less symmetric cases in sect.~\ref{Summary}.

So, we are led to consider the most general low-energy theory for three scalar fields obeying Poincar\'e invariance and the internal symmetries \eqref{shift} and \eqref{rot}. The shift \eqref{shift} forces each field to appear with at least one derivative. Lorentz invariance forces these derivatives to contract among themselves. At lowest order in the derivative expansion, the only Lorentz-scalar, shift-invariant quantity is the matrix
\be \label{BIJ}
B^{IJ} \equiv \di_\mu \phi^I \, \di^\mu \phi^J \; .
\ee
We then have to construct $SO(3)$ invariants out of this matrix. For a $3 \times 3$ matrix, there are only three independent ones, which we can take for instance to be the traces
\be
[B] \; , \quad [B^2] \; , \quad [B^3] \; ,
\ee
where the brackets $[\dots]$ are shorthand for the trace of the matrix within. Alternatively, one could take the determinant, and two of the traces above. In the following, we will find it convenient to use one invariant---say $[B]$---to keep track of the overall `size' of the matrix $B$, and to choose the other two such that they are insensitive to an overall rescaling of $B$, e.g.
\be
X = [B] \;, \qquad Y = \frac{[B^2]}{[B]^2} \;, \qquad Z = \frac{[B^3]}{[B]^3} \; .
\ee
The most general solid action therefore is
\be \label{solid action}
S = \int \! d^4 x \, F\big( X,Y,Z \big )  + \dots
\ee
where $F$ is a generic function that depends on the physical properties of the solid---e.g.~its equation of state---and the dots stand for higher-derivative terms, which are negligible at low energies and momenta.

As a self-consistency check, notice that the background configuration \eqref{vev} solves the equations of motion for any value of $\alpha$, which can then be thought of as being determined by the boundary conditions (the external pressure we alluded to above.)
The eom for the action above are
\be
\di^\mu \Big( \di_\mu \phi^J  \frac{\di F}{\di B^{IJ} }\Big) = 0 \; .
\ee
For a linear configuration like \eqref{vev}, all terms in parentheses are constant, since they depend on the $\phi^I$s' first derivatives, which are constant. The eom are thus trivially obeyed.

As a side note, a perfect fluid fits into the same general description. Only, it features infinitely many more internal symmetries, which express the physical fact that for a fluid one can move volume elements around in an adiabatic manner without paying any energy price---only their compression matters. By contrast, if one tries to displace a solid's volume elements, even without compressing or dilating them, one encounters stresses that want to bring the volume elements back to their rest positions. Mathematically, in our language this property of a fluid is expressed as the invariance of the dynamics under internal volume-preserving diffeomorphisms:
\be \label{diff}
\phi^I \to \xi^I(\phi) \; , \qquad \det \frac{\di \xi^I}{\di \phi^J} = 1 \; .
\ee
Of the $SO(3)$ invariants above, only one particular combination survives: the determinant of $B$,
\be \label{detB}
\det B = \sfrac16 \big( [B]^3 - 3 [B] [B^2] + 2[B^3] \big) \; ,
\ee
which is of course insensitive to multiplication of $B$ by unit-determinant Jacobians. At lowest order in the derivative expansion, a fluid's action is thus
\be \label{fluid action}
S _{\rm fluid} = \int \!  d^4 x \,  F\big( \det B \big) + \dots
\ee
So, in this formalism, a fluid is just a very symmetric solid. 

Back to the solid. The background configurations \eqref{vev} spontaneously break some of our symmetries. There are associated Goldsone bosons, which are nothing but fluctuations of the $\phi^I$'s about such a background,
\be
\phi^ I = \alpha (x^I + \pi^I) \; .
\ee
We get these fluctuations' free action by expanding our action \eqref{solid action} to second order in $\pi^I$. Using
\be
B^{IJ} = \alpha^2 \big( \delta^{IJ} + \di^I \pi^J + \di^J \pi^I + \di_\mu \pi^I \di^\mu \pi^J  \big) \; , 
\ee
after integrating by parts and neglecting boundary terms we get
\begin{align}
S \to S_2 & = \int \! d^4 x \, \Big[-\sfrac13  F_X X \cdot \dot{\vec \pi} \,^2 +\big( \sfrac13 F_X X +\sfrac{6}{27} (F_Y+F_Z) \big) \, (\di_i \pi_j )^2   \nonumber \\
& + \big( \sfrac19 F_{XX} X^2 +\sfrac{2}{27} (F_Y+F_Z) \big) \, (\vec \nabla \cdot \vec \pi )^2 \Big] \; ,
\label{quadratic}
\end{align}
where the subscripts stand for partial derivatives, which are to be evaluated at the background values
\be
X \to 3 \alpha^2 \; , \qquad Y \to 1/3 \; , \qquad Z \to 1/9 \; .
\ee
These Goldstone excitations are the solid's phonons. For what follows it will be convenient to split the phonon field $\vec \pi$ into a longitudinal  part and a transverse  one,
\be
\label{decomposing pi}
\vec \pi = \vec \pi_L  + \vec \pi_T \; , \qquad \quad \vec \nabla \times \vec \pi_L = 0 \; , 
\quad \ \vec \nabla \cdot \vec \pi_T = 0\; .
\ee
It is straightforward to extract the longitudinal and transverse propagation speeds from the phonon's action:
\be \label{cLcT}
c_L^2 = 1 +  \frac23 \frac{F_{XX} X^2}{F_X X} + \frac8{9} \frac{(F_Y + F_Z)}{F_X X} \; , \qquad c_T^2 = 1 + \frac23 \frac{(F_Y + F_Z)}{F_X X}  \; ,
\ee
in terms of which the quadratic action is simply
\be \label{quadratic2}
S_2  = \int \! d^4 x \, (-\sfrac13 F_X X) \, \Big[ \dot{\vec \pi} \,^2 
- c_T^2 \, (\di_i \pi_j )^2  -  (c_L^2-c_T^2) \, (\vec \nabla \cdot \vec \pi )^2  \Big] \; .
\ee

If we expand eq.~\eqref{solid action} to higher orders, we get the interactions among the phonons. The expansion is straightforward, but already at cubic order the result is quite messy, and not particularly illuminating.
Below, we will display explicitly the cubic interactions in a particular limit, which yields some simplifications, and which is the physically relevant limit for an inflationary background. For the moment, it suffices to say that, by construction, the $n$-th order interaction terms will be schematically of the form $(\di \pi)^n$, where the derivatives can be spacial or temporal, and the indices are contracted in all possible ways. The coefficients of these interactions terms---the coupling constants---will be given by suitable derivatives of $F$, evaluated on the background solution. Like for all derivatively-coupled theories, our interactions become strong in the UV, at some energy scale  $\Lambda_{\rm strong}$. For our theory to be predictive for cosmological observables, we will need this strong-coupling scale to be above the Hubble rate $H$, for the whole duration of inflation.

%%%%%%%%%%%%%%%%%%%%%%%%%%%%%%%%%%
%%%%%%%%%%%%%%%%%%%%%%%%%%%%%%%%%%
\section{Inflation}\label{inflation}

We can now allow for a cosmological spacetime metric and for dynamical gravity, which, operationally, is trivial: the index-contraction in \eqref{BIJ} should be done via $g^{\mu\nu}$ rather than $\eta^{\mu\nu}$, and the measure in \eqref{solid action} should carry a $\sqrt{-g}$. 
As usual, `minimal coupling' corresponds to the most general coupling one can have between a matter system and gravity at lowest order in the derivative expansion.
Then our solid's stress-energy tensor is
\be \label{Tmn}
T_{\mu\nu} = -\frac{2}{\sqrt{-g}} \frac{\delta S}{\delta g^{\mu\nu}} = - 2 \, \frac{\di F}{\di B^{IJ}} \, \di_\mu \phi^I \di_\nu \phi^J + g_{\mu\nu} \, F \; .
\ee

As to the scalar fields' background configuration, the $x^I$ in \eqref{vev} should now be interpreted as {\em comoving} FRW coordinates. The reason is that the FRW metric is invariant under translations and rotations acting on the comoving coordinates, and we want the l.h.s.~and the r.h.s.~in \eqref{vev} to transform in the same way under the  symmetries we are trying to preserve. We can also choose the normalization of the comoving coordinates to set the $\alpha$ parameter to one,  so that from now on the background configuration is simply
\be
\langle \phi^I \rangle = x^I \; .
\ee

When computed on the background, the stress-energy tensor reduces to the standard $T^{\mu} {}_\nu= {\rm diag} \big( -\rho, p , p , p \big)$, with
\be \label{rho and p}
\rho= - F \; , \qquad p = F - \sfrac2{a^2} F_X  \; ,
\ee
where the subscript $X$ stands for partial derivative, and $F$ and $F_X$ are evaluated at the background values for our invariants:
\be \label{XYZ background}
X \to 3 /{a^2(t)} \; , \qquad Y \to 1/3 \; , \qquad Z \to 1/9 \; .
\ee 
Notice that---by construction---$X$ is the only invariant that depends on the scale factor; $Y$ and $Z$ were designed to be insensitive to an overall rescaling of $B^{IJ}$. This is the reason why only $F_X$  appears in the pressure: for an FRW solution, the pressure is related to the response of the system to changing the scale factor, i.e., the volume. For a more general configuration, the stress-energy tensor \eqref{Tmn} has a more complicated structure, which depends on $F_Y$ and $F_Z$ as well, which we report here for later use:
\be
\label{Stress Tensor}
T_{\mu\nu}=g_{\mu\nu}\,F -2\,\pd_\mu \phi^I \pd_\nu \phi^J \left(\Big(F_X-\frac{2F_Y Y}{X}-\frac{3F_Z Z}{X} \, \Big)\delta^{IJ}+\frac{2 F_Y B^{IJ}}{X^2}+\frac{3F_Z B^{IK} B^{KJ} }{X^3}\right) \; .
\ee

Now, in order to have near exponential inflation, we need
\be \label{slow roll}
\epsilon \equiv -\frac{\dot H}{H^2} \ll 1 \; .
\ee
Via the Friedmann equations
\footnote{We are defining the Planck scale as $\mpl^2 = (8\pi G)^{-1}$}, 
\be
H^2 = \frac{1}{3\mpl^2} \rho \; , \qquad \dot H = - \frac{1}{2\mpl^2} (\rho+p) \; ,\label{Friedman}
\ee
and eq.~\eqref{rho and  p}, we can express $\epsilon$ directly in terms of our Lagrangian $F$:
\be \label{epsilon}
\epsilon = 3 \cdot \frac{1}{a^2}\frac{F_X}{F} =  \frac{\di \log F}{\di \log X} \; ,
\ee
where we used eq.~\eqref{XYZ background} for the background value of $X$.
We thus see that if we want our solid to drive near exponential inflation, we need a very weak $X$-dependence for $F$. Which is not surprising, since $X$ is the only invariant that is sensitive to the volume of the universe: for inflation to happen, the solid's energy should not change much if we dilate the solid by $\sim e^{60}$; this is only possible if the solid's dynamics do not depend much on $X$. 

This also suggests how to enforce the smallness of $\epsilon$ via an approximate symmetry. Consider the scale transformation
\be \label{scale}
\phi^I \to \lambda \, \phi^I \;, \qquad \lambda= {\rm const} \; .
\ee
The matrix $B^{IJ}$ changes by an overall $\lambda^2$ factor, which affects $X$ but not $Y$ nor $Z$. Therefore, the smallness of $F_X$ can be interpreted as an approximate invariance under \eqref{scale}: If $F$ only depended on $Y$ and $Z$, it would be exactly invariant under \eqref{scale}, and this would prevent quantum corrections from generating some $X$-dependence. If we start with a small $F_X$ at tree level, the symmetry is only approximate, yet all further $X$-dependence generated at quantum level will be suppressed by the small symmetry-breaking coupling constant---$F_X$ itself.

Notice that in general, {\em ordinary} scale invariance---that acting on the spacetime coordinates as $x^\mu \to \lambda x^\mu$---cannot be readily used as an approximate symmetry to enforce the smallness of symmetry-breaking parameters. For instance, one cannot solve the Higgs-mass hierarchy problem this way \cite{RZ}. Moreover, it is generically anomalous, that is, broken by quantum effects, as clearly displayed by the running of coupling constants for interactions that are scale-invariant at tree level. Here instead, we are dealing with a purely internal symmetry---eq.~\eqref{scale}---which commutes with all spacetime symmetries. It has nothing to do with spacetime scale-invariance. It is on an equal  footing with our other internal symmetries \eqref{shift}, \eqref{rot}, and, like those, is non-anomalous and {\em can} be used to constrain the structure of the Lagrangian. To avoid confusion, in the following we will refer to the symmetry \eqref{scale} as `internal scale invariance'.

Notice in passing that at lowest order in the derivative expansion, imposing internal scale invariance gives us full (internal) conformal invariance as a byproduct. The reason is the following. The invariants we are using in our lowest-order action are combinations of traces of the form $[B^n]$. Under a generic internal diff
\be
\phi^I \to \xi^I(\phi) \; ,
\ee 
these traces transform as
\be
[B^n] \to [(J^T  J) \cdot B \cdot (J^T   J) \cdot B \cdots (J^T   J) \cdot B  ] \; ,
\ee
where $J$ is the diff's Jacobian matrix, and we have used the cyclicity of the trace. Now, {\em by definition}, conformal transformations are the subgroup of diffs that change the flat metric $\delta^{IJ}$ only by an overall scalar factor, that is, they have a $(J^T  J)$ proportional to the identity. So, under a general conformal transformation, our traces only change by an overall ($\phi$-dependent) factor. Our $Y$ and $Z$ combinations are insensitive to such a change, and as long as the Lagrangian only depends on those, it is invariant under all internal conformal transformations. It is interesting that even though our internal scale invariance has in principle nothing to do with ordinary spacetime scale invariance, it shares with it the unavoidable company of special conformal transformations. In both cases, special conformal transformations are not needed to close the symmetry group, yet they are respected by generic scale-invariant dynamics. In the spacetime symmetry case, there are fundamental reasons why that happens \cite{Polchinski, LPR}. In our case, it appears to be an accidental feature of the lowest order truncation of the derivative expansion.

Back to physics. The smallness of $\epsilon$ in our model does not come for free. To see this, notice  that by  dialing $F_X$ we could go continuously through $\epsilon = 0$ and end up with negative $\epsilon$, that is, positive $\dot H$, which violates the null energy condition (NEC). But our EFT conforms to the hypotheses of the general theorem of \cite{DGNR}, which links NEC violations to such pathologies as ghost- or gradient-instabilities and superluminality. For small but positive $\epsilon$, we could be dangerously close to these pathologies
\footnote{These considerations are irrelevant for standard slow-roll inflation, where the smallness of $\epsilon$ is achieved via an approximately flat potential. One cannot play with the potential's slope and end up with positive $\dot H$. Positive $\dot H$ would require flipping the sign of the inflaton's kinetic energy, which would of course entail ghost-instabilities. On the other hand, in our case the sign of $\dot H$ is the same as that of $F_X$, which is the Lagrangian parameter we are playing with to make $\epsilon$ small.}.  Our quadratic action for the phonons---eq.~\eqref{quadratic}---is quite explicit about this: the phonon's kinetic energy is suppressed by $\epsilon$,
\be \label{kinetic}
S_2 \sim \int \! d^4 x \,   \epsilon | F| \cdot  \dot{\vec \pi} \, ^2 + \dots
\ee
For very small $\epsilon$, this can in principle lead to two problems for our theory:
\begin{itemize} 
\item
Superluminality: the gradient energies in \eqref{quadratic} are {\em not} explicitly suppressed by $\epsilon$, and as a consequence the propagation speeds \eqref{cLcT} are formally of order $1/\epsilon$, unless the numerators are also small. In an effective field theory like ours, with spontaneously broken Lorentz invariance, superluminal signal propagation is not necessarily an inconsistency. However, it prevents the theory from admitting a standard Lorentz-invariant UV-completion \cite{AADNR}. We therefore feel that it should be avoided.
\item
Strong coupling: unless interactions are also suppressed by suitable powers of $\epsilon$---and it turns that they are not---a smaller kinetic energy means stronger interactions. This is obvious if one goes to canonical normalization for $\pi$, by absorbing the prefactor in \eqref{kinetic} into a redefined phonon field. Then inverse powers of $\epsilon$ will show up in the interaction terms, thus signaling that the strong coupling scale of the theory is suppressed by some (positive) power of $\epsilon$. We have to make sure that this strong coupling scale is above $H$, for the whole duration of inflation.
\end{itemize}

As for the former issue, notice first of all that the term proportional to $F_{XX}$ in the expression for $c_L^2$ is forced to be close to $-2/3$. The reason is that not only do we need the `slow-roll' condition \eqref{slow roll} for inflation to happen, we also need 
\be \label{eta}
\eta = \frac{\dot \epsilon}{\epsilon H} \ll 1
\ee
for inflation to last many $e$-folds.\footnote{In the  computations that follow, we will assume all the slow-roll parameters to be of the same order of magnitude.} This forces the second derivative $F_{XX}$ also to be small. In particular, given eq.~\eqref{epsilon}, and
\be
H = \frac{d}{dt} \log a = - \frac12 \frac{d}{dt} \log X \; ,
\ee
we get
\be
\frac{F_{XX}X^2}{F_X X} = -1 +\epsilon - \sfrac12 \eta \; .
\ee
So that, at lowest order in $\epsilon$ and $\eta$, the propagation speeds \eqref{cLcT} reduce to
\be
c_L^2 \simeq \frac13 + \frac8{9} \frac{(F_Y + F_Z)}{F_X X} \; , \qquad c_T^2 = 1 + \frac23 \frac{(F_Y + F_Z)}{F_X X} \; .
\ee
It is quite interesting that in this limit they depend on exactly the same $(F_Y+F_Z)$ combination; we see no obvious reason why this should be the case. As a result, the two  speeds are not independent: they are related by
\footnote{To all orders in $\epsilon$ and $\eta$, the exact relation is 
\be
c_T^2 = \sfrac34\big (1+ c_L^2 - \sfrac23 \epsilon +\sfrac13 \eta \big) \; . \label{cVTrelation}
\ee}
\be \label{cT}
c_T^2 \simeq \sfrac34(1+ c_L^2) \; .
\ee
We thus see that for both speeds to be sub-luminal, we need 
\be
c_L^2 < \sfrac13 \; ,
\ee
that is, positive $(F_Y+F_Z)$ (recall that $F_X X = \epsilon F$ is negative, because $F$ is). We do not want $(F_Y+F_Z)$ to be too positive though---otherwise, we end up with negative squared speeds, that is, exponentially growing modes. We thus need $(F_Y+F_Z) $ to fit into a small window,
\be \label{window}
0 < (F_Y+F_Z) < \sfrac38 \epsilon |F| \; .
\ee
We were able to motivate the smallness of $F_X$,
\be
F_X X = \epsilon F \; ,
\ee
via an approximate symmetry. It would be desirable to do the same for this new combination of derivatives. One possibility is of course to say that {\em all} derivatives of $F$ are small, that is
\be
\frac{\di F}{\di B^{IJ} } B^{KL} \sim \epsilon F \; .
\ee
This formally corresponds to an approximate invariance under {\em all} internal diffs
\be
\phi^I \to \xi^I(\phi) \; ,
\ee
that is, to the  statement that the value of the Lagrangian does not depend much on the fields. Less formally, and more physically, it corresponds to saying that the bulk of the solid's energy density and pressure are dominated by a cosmological constant, which does not depend on the fields. Although this is of course a technically natural choice---having a {\em large} cosmological constant was never a problem---it is not particularly interesting. It would be more interesting to find a symmetry that allows large derivatives of $F$,
\be \label{largeFY}
F_Y, F_Z \sim F
\ee
but that---in the limit of exact symmetry---forces the combination $F_Y + F_Z$ to vanish. Another possibility would be a symmetry that makes $F_Y + F_Z$  saturate the {\em upper} bound in \eqref{window}, that is, that makes $c_L^2$ vanish. This is not as unlikely as it sounds: for instance the perfect fluid action \eqref{fluid action}---whose structure is protected by the volume-preserving diff symmetry---features vanishing propagation speed for the {\em transverse} phonons \cite{DGNR, ENRW}, as can be checked explicitly in \eqref{cLcT}, by using expression \eqref{detB} for the determinant. 
We have not been able to find a symmetry that enforces the condition \eqref{window} while preserving \eqref{largeFY}. We have to take such a condition as an assumption, which might involve some fine tuning, but which is nonetheless consistent and necessary for the consistency of our inflationary solution.
 
As for the strong coupling issue, we have to estimate the strong coupling scale $\Lambda_{\rm strong}$ in our small $\epsilon$ limit, and make sure that cosmological perturbations are weakly coupled at horizon crossing, that is, at frequencies of order $H$. 
Expanding the action \eqref{solid action} to all orders in $\pi$ we get interactions of the form
\be
f_n \cdot (\di \pi)^n \; ,
\ee
where $f_n$ is some typical derivative of $F$. In our case some combinations of derivatives are small,
\be
F_X X \sim (F_Y+ F_Z) \sim \epsilon F \; ,
\ee
but we do not expect this to yield a substantial weakening of interactions. For instance, we will see below that in our approximation the coefficient weighing cubic interactions is $F_Y$, which, as we argued above, can be as large as the background energy density, $F_Y \sim F$. Assuming that $F_Y$ is a good estimate for the coefficients encountered in interaction terms, and assuming for the moment that both $c_L$ and $c_T$ are of order of the speed of light---so that there is no parametric difference between time- and space-derivatives---we can estimate very easily the strong coupling scale: We can go to canonical normalization for the kinetic term
\be
{\cal L}_2 \sim \epsilon F \cdot (\di \pi)^2 \to (\di \pi_c)^2 \; ,
\ee
so that the $n$-th order interaction becomes
\be
{\cal L}_n \sim F_Y \cdot (\di \pi)^n  \to \frac{F_Y}{(\epsilon F)^{n/2}}  (\di \pi_c)^n \; .
\ee
This is a dimension-$2n$ interaction, weighed by a scale
\be
\Lambda_n \sim \Big(\frac{\epsilon^n F^n}{F_Y^2}\Big)^\frac1{4n-8}
\ee
(recall that $F$ and $F_Y$ have mass-dimension four.)
If $F_Y$ is of the same order as $F$, this is simply
\be
\Lambda_n \sim F^{1/4} \cdot \epsilon^\frac{n}{4n-8} \; , \qquad (F_Y \sim F) \; ,
\ee
which, for $n \ge 3$ and $\epsilon \ll 1$, is an increasing function of $n$. The lowest of all such scales---which defines the strong coupling scale of the theory---is thus that associated with $n=3$:
\be
\Lambda_{\rm strong} = \Lambda_3 \sim F^{1/4} \epsilon^{3/4} \; , \qquad (F_Y \sim F) \; .
\ee
If on the other hand $F_Y$ is much smaller that $F$, say of order $\epsilon F$, we get that all interactions are weighed by the same scale, which then defines strong coupling:
\be
\Lambda_{\rm strong} \sim \Lambda_n \sim F^{1/4} \cdot \epsilon^{1/4} \; , \qquad (F_Y \sim \epsilon F) \; .
\ee
Either way, the strong coupling scale is a fractional power of $\epsilon$ smaller than the scale associated with the solid's energy density.

If the propagation speeds $c_L, c_T$ are non-relativistic, the estimate of the strong coupling scale depends on the specific structure of the interaction terms, that is, on how many time-derivatives there are. In general, one may expect stronger interactions, i.e., lower strong-coupling scales for non-relativistic excitations (see e.g.~\cite{ENRW} for a systematic analysis of this phenomenon in a different limit of our solid action.)
Notice first of all that, because of \eqref{cT}, the transverse phonon speed $c_T$ is always relativistic,
\be
\label{cT is rel}
\sfrac34 < c_T^2 < 1 \; . 
\ee
So, our estimates above always work for the transverse phonons' self-interactions. For longitudinal phonons with $c_L \ll 1$, we can repeat the estimate using the cubic interaction, assuming this is still a good indicator of the strong coupling scale of the theory.
Expanding \eqref{solid action} up to cubic order, and neglecting terms that are proportional to $X$-derivatives of $F $ or to $(F_{Y}+F_Z)$, we find
\begin{align}
S_3 \simeq \int \! d^4 x \, \big(-\sfrac{1}{243} F_Y \big) \cdot \Big\{ 16 \, [\di\pi]^3 -  
36 \, [\di \pi]^2 \, \big( [\di \pi \cdot \di \pi^T] +  [(\di \pi)^2] \big)  \nonumber \\
+18 \, [(\di \pi)^3] + 18 \, [(\di \pi)^2 \cdot \di \pi^T]\Big\} \; ,  \label{cubic}
\end{align}
where $(\di \pi)_{ij} \equiv \di_i \pi_j$ is the matrix of {spacial} derivatives of $\pi$, $\di \pi^T$ is its transpose, and the brackets stand for the trace.
Notice in particular that there are no time-derivatives. For $c_L \ll 1$, one can estimate the strong coupling scale via the following trick \cite{ENRW}. We can redefine the time variable as $t \to t'/c_L$. Now in the kinetic energy term there is no hierarchy between time- and space-derivatives,
\be
S_2 \sim \int d^4 x \, \epsilon F \cdot \big( \dot \pi^2 - c_L^2 (\nabla \pi)^2 \big) \to \int d^4 x' \, \epsilon F \, c_L \cdot  (\di ' \pi)^2 \; ,
\ee
and we can apply the usual order-of-magnitude estimates as for relativistic theories. The cubic interaction becomes schematically
\be
S_3 \to \int d^4 x' \, \frac{F_Y}{c_L} \cdot  (\di ' \pi)^3 \; . 
\ee
To get the lowest possible value for the strong-coupling scale---that is the most dangerous one---we can take $F_Y$ to be as large as possible,  $F_Y \sim F$.
\footnote{There is a third possibility where $F_{Y} \gg F$. While this may not be a problem when analyzing the strong coupling scale of the theory, it is still very unnatural to have the background energy scale much smaller than the other scales in the theory. We already have one cosmological constant problem in cosmology, it would maybe be better if we tried not to introduce a second one.}
Going to canonical normalization and estimating the strong-coupling scale as above we get
\footnote{If one were to repeat the same analysis for a more generic $n$-th order interaction, also weighed by $F_Y$ like eq.~\eqref{cubic}, and also involving spatial derivatives only, one would get
\be
p_{\rm strong } \sim F^{1/4} \epsilon^\frac{n}{4(n-2)} c_L^\frac{n+2}{4(n-2)} \; ,
\ee
which, for small $\epsilon$ and $c_L$, is minimized at $n=3$.}:
\be
p_{\rm strong } \sim F^{1/4} (\epsilon^3 c_L^5)^{1/4}
\ee
This is the strong coupling {\em momentum} scale, or equivalently, the strong-coupling energy scale in  units that are appropriate for our new $t'$ variable. To convert to the original units of energy, we have to multiply by an extra $c_L$:
\be
E_{\rm strong } \sim F^{1/4} (\epsilon^3 c_L^9)^{1/4} \; .
\ee

As we mentioned above, cosmological perturbation theory is under control only if
\be 
E_{\rm strong } \gg H \; .
\ee
Relating $H$ and $F$ via the Friedmann equation, $H^2 \sim F/\mpl^2$, we get a lower bound on the combination $\epsilon \cdot c_L^3$,
\be \label{epsilon c^3}
\epsilon \cdot c_L^3 \gg \big( H/ \mpl \big)^{2/3} \; .
\ee
In principle our $H$ can be several orders of magnitudes smaller than the Planck scale, in which case this bound is not particularly restrictive.
Still, it is a nontrivial condition for the self-consistency of the perturbative computations we will perform. 

Once the strong-coupling danger is exorcised, large interactions are demoted (or promoted) from a problem to an exciting feature of the model: they imply huge non-gaussianities for our cosmological perturbations. As we will see in sect.~\ref{perturb}, our non-gaussian signal is peaked on squeezed triangles, with the same size-dependence as the so-called local forms of non-gaussianity, but with a different angular dependence. The corresponding $f_{\rm NL}$ parameter is of order $1/ (\epsilon \cdot c_L^2)$, which is a factor of $1/\epsilon$ bigger than what one finds---at the same value of $c_L$---in single-field models with non-relativistic sound speed.

A clarification is in order: we have been analyzing the viability of our model focusing on the phonons' dynamics, neglecting the background spacetime curvarture and the phonons' mixing with gravitational perturbations. Of course this is not entirely correct. As we mentioned in the Introduction however, at energies much bigger than $H$, or equivalently, for time-scales much shorter than $H^{-1}$, curvature and mixing have negligible effects, and in first approximation they can be neglected. Our conditions above, \eqref{window} and \eqref{epsilon c^3}, should then be thought of as necessary and sufficient for our system to be well-behaved in the UV, at very short distances and time scales. Our detailed analysis of cosmological perturbations in sect.~\ref{perturb} will confirm these results.

We should also point out that although we will be using standard `slow-roll' nomenclature for the conditions \eqref{slow roll}, \eqref{eta} and for the associated perturbative expansion, nothing is `rolling' in our system, slowly or otherwise: our $\phi^I$ scalars are {\em exactly} constant in time. As usual however, the so-called slow-roll expansion really relies on the slowness of certain time-dependent observables like $H$, $\dot H$, etc., which are well defined regardless of the presence of a rolling scalar. We will still use `slow-roll' to refer to such a weak time-dependence, hoping that this will not cause confusion. As we emphasized, in our case the physical origin of this slowness is the near independence of the dynamics on $X$, which is, among our invariants, the only one that depends on time. Besides $\epsilon$ and $\eta$, in the following we will need one more slow-roll parameter,
\be
\label{s}
s \equiv \frac{\dot c_L}{c_L H} \; ,
\ee
which is small, because $c_L$ depends on time only via the Lagrangian's $X$-dependence.

Finally, we should comment on why we are focusing on a solid rather than on a perfect fluid. First, since eventually we will be interested in quantum mechanical effects---as usual, quantum fluctuations will be the `seed' for cosmological perturbations---we focus on a solid because we do not know yet how to consistently treat the perfect fluid effective theory as a quantum theory \cite{ENRW}.
The problem has to do with the transverse excitations, which appear to be strongly coupled at all scales. Second, even forgetting about the transverse excitations and focusing on the longitudinal ones, we would not be able to keep those weakly coupled for many $e$-folds.
As clear from \eqref{cVTrelation}, to have vanishing $c_T^2$ (which is one of the defining features of a fluid) and small $\epsilon$, we need $\eta \sim -1$. But, by definition, $\eta = \dot \epsilon/(H \epsilon)$, so that we need $\epsilon = F_X X /F$ to decrease by an order one factor over one Hubble time, i.e.~to decrease like some order-one power of $1/a$. $F$ has to be nearly constant over many $e$-folds, which means that it is actually the numerator $F_X X $ that is tracking $1/a(t)$. But it is precisely combinations like $F_X X$ that control the strong-coupling scale for longitudinal excitations in a fluid \cite{ENRW}, which means that we cannot have $F_X X$ decrease by exponentially large factors  without making the system strongly coupled at frequencies of order $H$ at some point during inflation.

%%%%%%%%%%%%%%%%%%%%%%%%%%%%%%%%%%
%%%%%%%%%%%%%%%%%%%%%%%%%%%%%%%%%%
\section{Physical `clocks' and reheating}\label{clock}

Eventually we want our inflation to end and to be followed by a standard hot Big-Bang phase, that is, we want the universe to reheat and to become radiation dominated
\footnote{It should be noted that \cite{Gruzinov} avoided going through a specific model of reheating by simply evolving the solid until $c_T\rightarrow0$, and demanding that the solid turn into a perfect fluid at that point. Given our analysis above, if we want to keep all the slow roll parameters small, as is reflected in (\ref{cT is rel}) $c_T^2$ {cannot} become much smaller than $3/4$. Via our EFT approach we thus see that the reheating model of \cite{Gruzinov} necessarily entails a breakdown of the slow-roll expansion before reheating, which is, of course, a consistent possibility.}. 
In our case, this process can be thought of as a phase-transition from a solid state to a relativistic fluid state. The advantage of our language in dealing with such a transition is that, as we emphasized in sect.~\ref{solids}, it describes both solids and fluids in terms of the same long-distance degrees of freedom, our scalars $\phi^I$. Only, the fluid action \eqref{fluid action} enjoys (many) more symmetries. So, regardless of the microscopic dynamics that are actually responsible for the phase transition, at long distances and time scales reheating corresponds to some sort of symmetry enhancement of our action. We will be more specific about this in a moment.

In terms of our infrared degrees of freedom, what triggers reheating? In standard slow-roll inflation, it is the inflaton  itself, when its time-dependent background field reaches a critical value. On the other hand,
in the absence of perturbations, our $\phi^I$'s are exactly time-independent: $\langle \phi^I \rangle = x^I$. However the metric is not, and there are  gauge-invariant combinations like our 
\be
X = g^{\mu\nu} \, \di_\mu \phi^I \di_\nu \phi^I \; ,
\ee
or the energy density and pressure in eq.~\eqref{rho and p}, that do depend on time. 
Usually we are used to solids turning into liquids---that is, melting---when the temperature exceeds a critical value.
But we can also envisage a solid that `melts' at zero temperature, when one of the physical quantities above goes past a critical value. Helium offers an example of such a  phenomenon: at zero temperature one can turn liquid helium into a solid  by raising the pressure beyond $\sim 25$ atm, and melt it back again by lowering the pressure below that value. In our case, we need this zero-temperature melting to be associated with a substantial release of latent heat, so that the fluid we end up with is (very) hot
\footnote{In the fluid phase, which is described by the action \eqref{fluid action}, the temperature is given by $T = - \frac{dF}{d \sqrt{\det B}}$ \cite{DHNS}.}.
As far as we can tell, this does not violate any sacred principles of thermodynamics.

Notice that as far as the background solution is concerned, to lowest order in the derivative expansion all choices for what observable triggers our solid's melting are physically equivalent: this is evident in our parameterization of the action \eqref{solid action}, where all time-dependent observables depend on time only through $X$, or, equivalently, through $a(t)$.
However, when we include fluctuations, we break this equivalence. For instance, the three invariants $[B]$, $[B^2]$, and $[B^3]$ are independent combinations of the fields' derivatives. In the presence of fluctuations, the hypersurface defined by $[B]$ reaching its critical reheating value is different from that defined by $[B^2]$ or $[B^3]$ reaching their critical reheating values. As a result, some of our predictions for cosmological correlation functions might depend on the physical variable chosen to trigger reheating. Notice that {\em after} reheating, in the hot fluid phase, there is no ambiguity: the lowest-order action \eqref{fluid action} only depends on one variable, the determinant of $B^{IJ}$.
It is thus natural, although not obviously necessary, to postulate that reheating is triggered by the value of $\det B$.

So, in terms of our action, our assumption is that for large $\det B$ the action has the general structure \eqref{solid action}, whereas for $\det B$ below a critical value, the action has the more restricted form \eqref{fluid action}. In the space of our $X$, $Y$, $Z$ invariants, this corresponds to dividing up the space into two regions, where the action has different symmetries: eqs.~\eqref{shift}, \eqref{rot} for the former, eq.~\eqref{diff} for the latter. 
Moreover, the slow-roll condition $\dot H \ll H^2$ during inflation is protected by another (approximate) symmetry of the solid phase, eq~\eqref{scale}, which we want  to be maximally violated in the post-reheating fluid phase, which has $\dot H \sim H^2$.
Since renormalization is local in field space, the existence  of different regions with different symmetries is protected by precisely those symmetries, and is thus a consistent and natural assumption
\footnote{A similar mechanism is at work in ghost inflation \cite{ACMZ}, or in the EFT description of finite-temperature superfluids \cite{Nicolis}.}.
For illustrative purposes, consider for instance the following action:
\be \label{example}
{\cal L} = F(X,Y,Z) \propto \bigg\{ \begin{array}{lll}
- (\det B)^{\epsilon/3} \cdot f(Y,Z) & &\mbox{for } \det B > 1 \\
-(\det B)^{2/3} & &\mbox{for } \det B < 1 \; ,
\end{array}
\ee
where $f$ is a generic function that evaluates to one for  the background values $Y \to 1/3$, $Z \to 1/9$, and we suppressed an overall common factor, which defines the density at reheating. The `gluing' at $\det B = 1$ can be smoothed at will.
In the first regime, ${\cal L}$ describes a solid driving an inflationary phase with constant $\epsilon$ (for fixed $Y$ and $Z$, $\det B$ scales like $X^3$---hence the $\epsilon/3$ power). In the second regime, ${\cal L}$ describes an ultra-relativistic fluid, with $p =  \rho/3 \propto T^4$ \cite{ENRW, DHNS}. The two regimes have different internal symmetries, as discussed above, but they share the same degrees of freedom. 
The classical evolution of the background solutions and of perturbations can then be followed smoothly through the transition region, as the long wavelength degrees of freedom are the same all along, and the equations of motion are regular
\footnote{As mentioned above, one should refrain from performing quantum computations in the fluid phase \cite{ENRW}.
So, eq.~\eqref{example} should not be thought of as a quantum effective theory in the second regime. Still, since at reheating all relevant modes are well outside the horizon and, thanks to the usual reasons, can be treated as classical, we only need \eqref{example} to be a consistent classical field theory, which it is.}.

Notice that we have been implicitly assuming that reheating is instantaneous, that is, that our solid/fluid phase transition happens in a time interval that is much shorter than the Hubble scale, which is reasonable in principle, but not necessary. One can also consider much slower transitions, which in field space would correspond to replacing the sharp critical values we have been talking about for our observables, with much more continuous transition regions. All our physical considerations above apply unaltered. Since, as we will discuss, some of our predictions are potentially model-dependent, for what follows we need to assume a specific model for reheating. So, we will assume that reheating is fast, much faster than $H$, and that it is controlled by the value of $\det B$.

%%%%%%%%%%%%%%%%%%%%%%
%%%%%%%%%%%%%%%%%%%%%%
\subsection{Why not the EFT of inflation?}
Before turning to a detailed analysis of cosmological perturbations, we close this section by discussing why our model does not conform to the standard EFT of inflation. As we just saw, we do have physical `clocks', that is time-dependent background observables, so why can't we use the standard results for spontaneously broken time-translations? The reason is that these time-dependent observables depend on time only because the metric does. 

To see why this subtlety is important, consider first the dynamics of our system at very short distances---in the so-called decoupling limit---where the matter fluctuations decouple from the gravitational ones and the Goldstone boson language is appropriate. In first approximation, this limit corresponds formally to setting $G$ to zero. But without gravity, our background solution has no time-dependence whatsoever! All observables like density, pressure, etc.~are now exactly constant in time, and only {\em spacial} translations and rotations are broken by the background configuration $\langle \phi \rangle = x^I$. As a result, the Goldstone bosons, whose existence and properties have to be assessed in the decoupling limit---because as recalled in the Introduction, only in this limit does it make sense to talk about them---are those associated with {\em this} spontaneous symmetry breaking pattern, not with time-translations. 

It is not surprising then, that once we re-introduce gravity, the dynamics of cosmological perturbations at all scales are quite different than for the EFT of inflation. This is manifest in the so-called unitary gauge, where one chooses the time-variable according to a physical clock. In the standard case, that clock would be the inflaton, and choosing the equal-time surfaces to be the equal-inflaton surfaces automatically sets to zero the inflaton perturbations and makes the metric the only fluctuating field. In our case, if we choose one of our time-dependent observables---say $\det B$ or $[B]$---to define unitary gauge, we are still left with {\em matter} perturbations, because the background time-dependence that we are using to set the gauge is carried by the metric, not by matter fields. We can either include the matter perturbations explicitly in the Lagrangian terms that we write down in this gauge, or we can set them to zero, by choosing the {\em spacial} coordinates now so that $\phi^I = x^I$. This is a complete gauge-fixing---all spacetime coordinates have been unambiguously defined---and is of course quite a different gauge choice than the standard unitary gauge. In particular, it is inequivalent to choosing the so-called $\zeta$-gauge for spacial diffs.  Either way, the Lagrangian terms one would write down are quite different than for the standard EFT of inflation. We will go into the details of this new unitary gauge in sect.~\ref{gauge app}.

The presence of matter fluctuations in the `naive' unitary gauge cannot be taken as a sign that we are dealing with what would be called a multi-field model in the standard classification. First, because there is a gauge---our `improved' unitary gauge---in which all matter fluctuations are set to zero. Second, because our spectrum of cosmological perturbations only includes {\em one} scalar mode, as clear from the Goldstone quadratic action \eqref{quadratic2}. Furthemore, as we will see, this scalar mode is not adiabatic. In other words, in our system there are no adiabatic modes of fluctuation. This is yet another manifestation that we are dealing with a truly unconventional cosmological system. 

Notice that at the classical level,  a subtle, isolated  exception to all of the above is offered by a perfect fluid. On the one hand, in our language a perfect fluid is just a very symmetric solid. In particular, it features the same symmetry breaking pattern. On the other hand however, because of powerful conservation laws for vorticity, classically one can consistently set to zero the transverse excitations---the vortices---and be left with an EFT for the compressional modes only \cite{DGNR}. This admits a description in terms of a single scalar which spontaneously breaks {\em time}-translations---a $P(X)$ theory---to which the standard EFT-of-inflation construction {\em is} applicable
\footnote{Here we are using the standard notation of the community: $P$ is a generic function and $X$ here does not refer to the $[B]$ we have been considering throughout the text but rather $\di_\mu \psi \di^\mu \psi$, where $\psi$ is some scalar field whose background vev is $\left<\psi\right>=\bar \psi (t)$.}.
In particular, for a perfect fluid scalar cosmological perturbations are adiabatic.
Once quantum effects are taken into account however, transverse excitations cannot be neglected any longer. In fact, for a fluid they are not particularly well behaved quantum mechanically \cite{ENRW}, which is one of the reasons why we have been considering a more generic solid rather than the special perfect-fluid case.

As a technical aside,
we should also emphasize that in a gauge where the matter fields are unperturbed, $\phi^I = x^I$, our $B^{IJ}$ matrix reduces simply to $g^{IJ}$, and our Lagrangian thus becomes the sum of the Einstein-Hilbert action and of a particular function of $g^{IJ}$, that is, it reduces to a Lorentz-violating theory of massive gravity.  Theories like this have been studied in broad generality in \cite{Dubovsky}. The reader familiar with the EFT of inflation might wonder why we are not writing down directly the action for the perturbations $\delta g^{IJ}$ in this gauge---the analog of unitary gauge in that case---, instead of going through the (apparently) unnecessary burden of writing an action for the full fields, solving for the background solution, and then expanding the action in small perturbations. The technical reason is that, unlike $\delta g^{00}$ or $\delta K_{ij}$ for the EFT of inflation, our $\delta g^{IJ}$ does not transform covariantly under the residual diffs, which are just time diffs for us.
The reason is that $g^{IJ}$ does, but its background value, $\frac{1}{a^2(t)} \delta^{IJ}$ does not. It is then technically more convenient to write an action for the full $g^{IJ}$, which just amounts to writing an action for $B^{IJ}$, like we have done.

%%%%%%%%%%%%%%%%%%%%%%%%%%%%%%%%%%
%%%%%%%%%%%%%%%%%%%%%%%%%%%%%%%%%%
\section{Cosmological perturbations}\label{perturb}
 
The three sections that follow contain a technical analysis of cosmological perturbations.
Before skipping directly to sect.~\ref{The size and shape of non-gaussianities}, the reader uninterested in the details of the derivations should be aware of our results: the scalar tilt \eqref{scalar tilt}, the tensor-to-scalar ratio \eqref{scalar to tensor ratio}, the tensor tilt \eqref{tensor tilt}, and the three-point function of scalar perturbations \eqref{three-point function} (which is analyzed in some detail in sect.~\ref{The size and shape of non-gaussianities}).

% \subsection{Preliminaries}
As the background stress tensor takes the usual homogeneous and isotropic form represented by $T^{\mu} {}_\nu= {\rm diag} \big( -\rho, p , p , p \big)$, all the interesting repercussions of our peculiar symmetry breaking pattern lie in the dynamics of perturbations around the slow roll background. 
% As is well known, doing perturbation theory in general relativity holds various complications and subtleties. 
In order to best isolate the dynamical degrees of freedom of the gravitational field it is most convenient to work in the ADM parametrization of the metric:
\be
\label{ADM decomposition}
ds^2=-N^2\; dt^2+h_{ij} \left(dx^i+N^i \; dt \right) \left(dx^j+N^j \; dt \right) \;.
\ee
It is straightforward to check that the inverse metric $g^{\mu\nu}$ is given by
\be
g^{00}=-\frac{1}{N^2},\qquad g^{0i}=g^{i0}=\frac{N^i}{N^2},\qquad g^{ij}=h^{ij}-\frac{N^i N^j}{N^2} \; ,
\ee 
where $h^{ij}$ is the inverse {\it spatial} metric: $h^{ik}h_{kj}=\delta^i_j$.
%This form is particularly convenient because when inserted into the action one find that the there are no time derivatives on $N^i$ and $N$. Thus, when the action is varied with respect to these fields their equations of motion are simply constraint equations. That is, they are non-dynamical. 
For the background FRW metric $N=1$,  $N^i=0$, and $h_{ij} = a^2(t) \, \delta_{ij}$.

Following \cite{Maldacena} we can write the action as
\be
\label{ADM action}
S= \int \! {\rm d}^4 x \;N \sqrt{h}\;\left\{\sfrac12 {\mpl^2} \big[R^{(3)} + N^{-2}(E_{ij}E^{ij}-E^2) \big]+\, F(X,Y,Z)\right\}
\ee
where $R^{(3)}$ is the 3-dimensional Ricci scalar constructed out of $h_{ij}$ and $E_{ij}=N \, K_{ij}$, with $K_{ij}$ denoting the extrinsic curvature of equal-time hypersurfaces. The constraint equations given by varying (\ref{ADM action}) with respect to $N$ and $N^i$ are: 
\begin{align}
\label{N constraint}
0&=\sfrac12 {\mpl^2}  \left[R^{(3)}-N^{-2}(E_{ij}E^{ij}-E^2)\right]+F(X,Y,Z)+N\frac{\pd F(X,Y,Z)}{\pd N}\\
\label{N^i constraint}
0&= \sfrac12 {\mpl^2}  \nabla_i\left[ N^{-1}(E^i_j-\delta^i_j E)\right] +N\frac{\di F(X,Y,Z)}{\di N^j} \; .
\end{align}
The derivatives of $F$ with respect to $N$ and $N^j$ can be calculated easily by noting that our $B^{IJ}$ (and hence $X, Y, Z$) can be expressed in ADM variables as 
\be
B^{IJ}=-\frac{1}{N^2}\big(\dot{\phi}^I-N^k\pd_k \phi^I\big)\big(\dot{\phi}^J-N^k\pd_k \phi^J\big)+h^{km}\pd_k\phi^I\pd_m\phi^J \; .
\ee 

%\subsection{Expanding in perturbations}
\label{Expanding in perturbations}

It is usually convenient to work in a gauge where scalar perturbations are removed from the matter fields and appear only in the metric, as $g_{ij} = a^2(t) \,(1+2 \zeta) \delta_{ij} $, see e.g.~\cite{Maldacena}.
This possibility is not available to us: By using up the three {\em spatial} diffs, we can set the matter field perturbations  to zero, $\phi^I = x^I$, but then the spatial metric has an extra scalar mode, proportional to $\di_i \di_j \chi$, which we now cannot remove in the usual manner.
However, we are still free to use time diffs, but these at best can set the scalar in front of $\delta_{ij}$, not $\chi$, to zero. A more useful choice is to use the time diff to set a physical ``clock''---like those we discussed in the last section---to its unperturbed value. If this clock controls reheating, then reheating will happen at the same time for all observers in this gauge. We review this gauge choice, which we call `unitary', in Appendix \ref{gauge app}.

For the moment we find it more convenient to work in spatially flat slicing gauge (SFSG)---defined in Appendix \ref{gauge app}---where we can write the fluctuations about the FRW background as
\be
\phi^I =x^I+\pi^I\; , \quad h_{ij}=a(t)^2 \exp(\gamma_{ij}) \; , \quad  N=1+\delta N \; ,
\ee
where $\gamma_{ij}$ is transverse and traceless, i.e.
\be
\di_i \gamma_{ij}=\gamma_{ii}=0 \; .
\ee
We can also further split the $\pi^i$ and $N^i$ fields in terms of their longitudinal scalar and transverse vector components. We therefore write:
\be
\pi^i=\frac{\di_i}{\sqrt{-\nabla^2}} \pi_L+\pi_T^i \; , \quad \text{and }\;N^i=\frac{\di_i}{\sqrt{-\nabla^2}} N_L+N_T^i \; ,
\ee
where $\di_i \pi_T^i=\di_i N_T^i=0$. From now on we will stop differentiating between internal $I, J, \dots$ indices and spacial $i,j, \dots $ ones. The reason is that of the full original $SO(3)_{\rm spacetime} \times SO(3)_{\rm internal} $ symmetry, only the diagonal combination is preserved by the background $\phi^I = x^I$. $\pi^i $ and $N^i$ both transform as vectors under this unbroken $SO(3)$, and therefore they carry the same kind of index.

For our purposes here we are interested only in the leading non-gaussian behavior. Barring accidental cancellations, this can be captured by keeping terms that are cubic in the fluctuations. In order to reproduce these terms it turns out to be necessary to only know $N$ and $N^i$ to first order in the fluctuations
\footnote {This lucky fact is because the higher order terms in $N$ and $N^i$ will be multiplying the constraint equations. In particular: the third order term of  $N$ and $N^i$ multiplies the zeroth order constraint equations, and the second order the first order constraint equations \cite{Maldacena}. If we where, however, to try and generate the fourth order terms we would need $N$ and $N^i$ to second order.}.
From now on, we find it easier to work in spacial Fourier space, with our convention defined for any field $\xi(x)$ by:
\be
\xi(t,\vec x) = \int_{\vec k} e^{i \vec k \cdot \vec x} \, \tilde{\xi}(t,\vec k) \; , 
\qquad \int_{\vec k} \equiv \int \frac{d^3 k}{(2 \pi)^3} \; .
\ee
For convenience, however, we will drop the twiddle as which field variable we intend will be obvious from the arguments. And so, solving the constraint equations (\ref{N constraint}) and (\ref{N^i constraint}) to first order in fluctuations we have
\bea
\label{delta N}
\delta N(t,\vec{k})&=&-\frac{a^2 \dot{H}}{k H}\,\frac{\dot{\pi}_L-\dot{H} \pi_L/H}{1-3\dot{H}a^2/k^2} \\
\label{N_L}
N_L(t,\vec{k})&=&\frac{-3a^2\dot{H}\dot{\pi}_L/k^2+\dot{H}\pi_L/H}{1-3a^2\dot{H}/k^2} \\
\label{N_T}
N_T^i(t,\vec{k})&=&\frac{\dot{\pi}_T^i}{1-k^2/4a^2\dot{H}} \label{NT}
\eea
where the dot denotes a time-derivative. 

Now, plugging these solutions back into (\ref{ADM action}) will give us the correct action for the fluctuations up to cubic order. For instance, the trilinear solid action after mixing with gravity is contained in Appendix \ref{trilinear action} while the quadratic actions for the tensor, vector and scalar modes are contained in the next section. 

Now that we have the correct action for the perturbations in the presence of an inflating background we can compute correlation functions.
In the end, we are interested in the post-reheating correlation functions of curvature perturbations, parameterized by either of the gauge invariant (at linear-order) combinations
\be
{\cal R} =\frac{A}{2}+H\delta u \; , \qquad {\cal \zeta} = \frac{A}{2}-H\frac{\delta \rho}{\rho}
\ee
where we have followed the notation of \cite{Weinberg}
\footnote{The general (i.e. before gauge fixing) perturbed metric (to the linear-order) is parametrized by 
\be
g_{ij}=a(t)^2\left(\delta_{ij}(1+A)+\pd_i\pd_j\chi + \pd_i C_j+\pd_j C_i + D_{ij}\right)
\ee
with $\pd_i C_i=0$ and $\pd_i D_{ij}=D_{ii}=0$; furthermore  the energy momentum tensor is decomposed into scalar, vector, and tensor modes as
\begin{align}
\delta T_{00}&=-\bar{\rho}\,\delta g_{00}+\delta \rho \label{T00}\\
\delta T_{i0}&=\bar{p}\,\delta g_{i0}-(\bar{\rho}+\bar{p})(\pd_i \delta u+\delta u_i^V) \label{Ti0}\\
\delta T_{ij}&=\bar{p}\,\delta g_{ij}+a^2(\delta_{ij} \delta p+\pd_i\pd_j \delta \sigma+\pd_i \delta \sigma^j+\pd_j \delta \sigma^i+\delta \sigma_{ij}^T) \label{Tij} \; .
\end{align}}.
During our solid inflation phase,  in spatially flat slicing gauge these are given by
%\subsection{Gauge Invariant Quantities}
\label{Gauge Invariant Quantities}
%
%Given our notation and the spacially flat slicing gauge, at linear order in perturbations, the gauge invariant  modes commonly used in the study of cosmology are given by
\be
\label{definition of adiabatic modes}
\mathcal{R}=-\frac{k}{3H \ep}\,\frac{\dot{\pi}_L+ H\ep\,\pi_L}{1+k^2/3a^2 H^2\ep} \; , \qquad \zeta=\sfrac{1}{3}\,\vec \nabla \cdot \vec \pi \;.
\ee
where the non-local piece of $\mathcal{R}$ comes from solving the constraint equation for $N_L$.

Two peculiarities concerning the behavior of these variables during solid inflation are worth mentioning at this point. First,  ${\cal R}$ and $\zeta$ do not coincide on super-horizon scales. Second, neither of them is conserved. These properties are in sharp contrast with what happens for adiabatic perturbations in standard cosmological models, and stem from the fact that during solid inflation, there are {\em no} adiabatic modes of fluctuation! We will clarify why this is the case in sect.~\ref{Why is zeta not conserved?}.

\section{Two-point functions}
\label{2ptfunction}
%\subsection{Quadratic Action}

Upon plugging the expressions \eqref{delta N}--\eqref{N_T} back into the action, the quadratic action
for tensor, vector, and scalar fluctuations reads:
\begin{align}
S^{(2)} &  =  S^{(2)}_\gamma+S^{(2)}_{T}+S^{(2)}_{L} \\
\label{graviton quadratic action}
S^{(2)}_\gamma &= \sfrac14 {\mpl^2} \int dt\,d^3x \,a^3\Big[\sfrac12 \dot{\gamma}_{ij}^2 -\sfrac{1}{2 a^2} \big(\pd_m \gamma_{ij}\big)^2 +2\dot{H}c_T^2 \, \gamma_{ij}^2 \Big]\\
\label{transverse quadratic action}
S^{(2)}_{T} &=\mpl^2 \int dt \int_{\vec k} \,a^3 \bigg[\frac{ k^2/4}{1-k^2/4a^2\dot{H}} \, \big| \dot{\pi}_T^i \big|^2 +\dot{H}c_T^2 \, k^2 \big|\pi_T^i  \big|^2 \bigg]\\
\label{longitudinal quadratic action}
S^{(2)}_{L} &= \mpl^2 \int dt \int_{\vec k}  \, a^3 \bigg[ \frac{ k^2/3}{1-k^2/3a^2\dot{H}}\big|\dot{\pi}_L -({\dot{H}}/{H})\pi_L\big|^2+\dot{H}c_L^2 \, k^2 \big| \pi_L  \big|^2 \bigg] \; .
\end{align}
Notice the quite nontrivial $k$-dependence for $S^{(2)}_{T}$ and $S^{(2)}_{L}$ in Fourier space, which would translate into a (spacially) non-local structure in real space.

\subsection{Tensor perturbations}

Using (\ref{graviton quadratic action}) we can calculate the two-point function of the tensor perturbations. As usual, it is a simpler calculation than the scalar case and will serve as a warmup.
We decompose the tensor modes into their polarizations
\be
\gamma_{ij}(\vec k,t)=\sum_{s=\pm} \epsilon_{ij}^s(\vec{k}) \gamma^s (\vec k,t) \;,
\ee
with $\epsilon_{ij}^{s}\epsilon_{ij}^{s'*}=2\delta^{ss'}$.
The transverse, traceless conditions on $\gamma_{ij}$ now simply become $\epsilon_{ii}=k_i \epsilon_{ij}=0$. We further decompose each $\gamma^s(\vec{k},t)$ as
\be
\gamma^s(\vec k, t)=\gamma^s_{cl}(\vec k, t) \; a^s(\vec k)+\gamma^s_{cl}(\vec k,t)^{*} \; a^{s\dag}(-\vec k) \;.
\ee
where $a^s(\vec k)^\dag$ and $a^s(\vec k)$ are creation and annihilation operators obeying the usual commutation relation
\be
[a^s(\vec k), a^{s'\dag}(\vec{k}')]=(2\pi)^3 \delta^{3}(\vec{k}-\vec{k}') \, \delta^{s s'}\;,
\ee 
and where the classical solution $\gamma^s_{cl}(\vec k, t)$ obeys the equations of motion  obtained by varying (\ref{graviton quadratic action}):
\be
\frac{d^2 }{d \tau^2}\gamma_{cl}+2aH \, \frac{d }{d \tau}\gamma_{cl}+\left(k^2+4\epsilon a^2 H^2 c_T^2\right)\gamma_{cl}=0 \label{tensoreom} \;.
\ee
In the above we have used conformal time $\tau$, where $d \tau=d t/a$, and the definition of the first slow-roll parameter (\ref{slow roll}). Using the time-dependence of $aH$, $\epsilon$, and $c_T$---which is worked out in Appendix \ref{time dep}---we can express the e.o.m. for the tensor mode (\ref{tensoreom}) up to first order in slow-roll parameters  as
\be
\frac{d^2 }{d \tau^2}\gamma_{cl}-\frac{2+2\ep_c}{\tau}\frac{d }{d\tau}\gamma_{cl}+\left(k^2+\frac{4\ep_c c_{T,c}^2}{\tau^2}\right)\gamma_{cl}=0 \;.
\ee
The subscript ``$c$'' denotes that the parameters $c_T^2,\ep$ are evaluated at some reference time $\tau_c$, which is chosen to be the (conformal) time when the {\it longest} mode of observational relevance today exits the horizon, i.e. $\tau_c$ is defined such that 
\be
\vert c_{L,c}k_{\rm min}\tau_c\vert \simeq \vert c_{L,c} \tau_c H_{\rm today}\vert=1
\ee
where the usual convention for flat spacetime $a_{\rm today}=1$ is understood.

The most general solution to the above equation takes the form
\be
\gamma_{cl}(\vec k,\tau)=(-\tau)^{3/2+\ep_c}\big[ \mathcal{A} \, H_{\nu_T}^{(1)}(-k\tau)+\mathcal{B} H_{\nu_T}^{(2)}(-k\tau) \big] \;, \qquad  \nu_T \simeq \sfrac{3}{2}+\epsilon_c -\sfrac{4}{3} c_{T,c}^2 \epsilon_c\;, \label{gsoltensor}
\ee
where $H^{(1,2)}$  are Hankel functions, and $\mathcal{A}$ and $\mathcal{B}$ are constants to be fixed by matching the appropriate initial conditions.

At very early times the {\it physical} wavelength---$k/a$---is so small compared to the Hubble scale $H$ that the curvature of spacetime cannot be perceived by such modes; it is therefore expected that the canonically normalized  classical solution should match the free wave function in the flat-space vacuum, $\frac1{\sqrt{2k}} e^{-i  k\tau}$.
Note that the canonically normalized field $\gamma^s_{\rm can.}(\vec k,\tau)$ is related to $\gamma^s(\vec k,\tau)$ by
\be
\gamma^s(\vec k, \tau)=\frac{\sqrt{2}}{\mpl} \frac{\gamma^s_{\rm can.}(\vec k,\tau)}{a(\tau)} \;.
\ee
Thus, the initial condition for $\gamma^s_{cl}(\vec k,\tau)$ is specified by
\begin{align}
\lim_{\tau \rightarrow - \infty} \gamma^s_{cl}(\vec k,\tau)=\frac{1}{\sqrt{k}\mpl a(\tau)}e^{-i k \tau} \;.
\end{align}
Comparing this to the general solution given by (\ref{gsoltensor}) and using the asymptotic form (for large arguments) of Hankel functions,
\be
\lim_{x\to +\infty}H^{(1)}_m(x)\to \sqrt\frac{2}{\pi x}e^{i x-i \left(m+\frac{1}{2}\right)\frac{\pi}{2}},\quad \lim_{x\to +\infty}H^{(2)}_m(x)\to \sqrt\frac{2}{\pi x}e^{-i x+i \left(m+\frac{1}{2}\right)\frac{\pi}{2}}
\ee
 it is enforced that
\be
\mathcal{A}=\frac{H_c(1-\ep_c)}{\mpl}\sqrt{\frac{\pi}{2}}\,(-\tau_c)^{-\ep_c} e^{i(\nu_T \pi/2+\pi/4)}+\mathcal{O}(\ep^2),\quad \mathcal{B}=0\;.
\ee
Where, once again, $H_c\equiv H(\tau_c)$ and $\ep_c=\ep(\tau_c)$. This result is valid up to first order in slow roll. 

It is interesting to note that even when these tensor modes are well outside the horizon, they are not conserved.  A similar story applies to the gauge invariant curvature perturbations $\zeta$ and ${\cal R}$ defined by (\ref{definition of adiabatic modes}) and the vector perturbation (like $\pi_T^i$), as we will see in the following section. This is in opposition to the usual situation in most inflation models, and will be discussed in more detail in Section \ref{Why is zeta not conserved?}. In particular,  by utilizing the asymptotic limit (for small argument) of the Hankel function
\be \label{asym Hankel small x}
\lim_{x\to 0^+}H_m ^{(1)}(x) \to (-i)\, \frac{\Gamma (m)}{\pi} \left(\frac{2}{x} \right)^m \; , 
\ee
the mild time-dependence of the tensor mode in late time is given by:
\be
\lim_{-k\tau\to 0^+}\gamma^s_{cl}(\vec k,\tau)=k^{-3/2}\left(\frac{\tau}{\tau_e}\right)^{4c_{T,e}^2\ep_e/3}(-k\tau_e)^{c_{L,e}^2\ep_e}\left(\frac{i H_e}{\mpl}+\mathcal{O}(\ep)\right) \;.\label{asym tensor}
\ee
where we have made use of relation (\ref{cT}). 
As we will see soon, the transverse vector modes and scalar modes in our model share this feature as well.

And so finally, we are ready to obtain the two-point function for the tensor perturbations of the metric. In particular, we are interested in its late time behavior, when modes are well outside the horizon:
\begin{align}
\big\langle\gamma^{s_1}(\vec k_1, \tau) \gamma^{s_2}(\vec k_2, \tau) \big\rangle&=  (2\pi)^3 \delta^{3}(\vec k_1+\vec k_2) \, \delta^{s_1 s_2} \, \big| \gamma_{cl}(\vec k_1, \tau)\big|^2  \\
&\stackrel{-k \tau \to 0^+}{\longrightarrow} (2\pi)^3 \delta^{3}(\vec k_1+\vec k_2) \, \delta^{s_1 s_2} \times 
\frac{H_c^2}{\mpl^2} \, \frac{1}{k_1^3} \, \frac{(\tau/\tau_c)^{8c_{T,c}^2\ep_c/3}}{(-k_1\tau_c)^{-2c_{L,c}^2\ep_c}}  \;.
\nonumber
\end{align}
The dependence on $k$ and $\tau$ is kept to first order in slow roll while the overall constant is to lowest order.  

The advantage of expressing time-dependent quantities in reference to a fixed fiducial time ($\tau_c$), as opposed to the usual convention of using the time at horizon crossing, is that the time- and momentum-dependence are made manifest. We can  simply read off the tilt of the spectrum to first order in slow roll from the above expression:
\be
\label{tensor tilt}
n_T-1 \simeq 2c_{L,c}^2\epsilon_c \;.
\ee
We can see that the two point function for tensor modes is {\em blue} shifted, which matches the result of \cite{Gruzinov}, and which is a distinctive signature of our scenario, unreproducible by more conventional models of inflation. As to the spectrum's overall amplitude, it is the usual one: $\langle \gamma \gamma \rangle \sim H^2/\mpl^2$.

\subsection{Scalar Perturbations}

We proceed by calculating the scalar two point function in a similar manner as above. As emphasized in section \ref{Gauge Invariant Quantities}, the scalar quantity of interest\footnote{We find $\zeta$ a more interesting quantity than $\mathcal{R}$ for the reason that given our assumption about reheating, $\zeta$ evolves continuously from inflation phase to post-inflation phase, while $\mathcal{R}$ does not. See Section \ref{reheating} for details. However, $\mathcal{R}$ does play a vital role as a simplifier in solving for the classical solution for scalar perturbation.  } is the gauge invariant quantity $\zeta$, which, in Fourier space, is related to the longitudinal Goldstones $\pi_L$ simply by $\zeta=-k\pi_L/3$ (see eq.~\eqref{definition of adiabatic modes}). 

As before, let's decompose the scalar field of interest in terms of creation and annihilation operators:
\begin{align}
\zeta(\vec k, t)&=\zeta_{cl}(\vec k, t) \; b(\vec k)+\zeta_{cl}(\vec k,t)^{*} \; b^{\dag}(-\vec k) \;,
\end{align}
where the usual commutation relation is obeyed $[b(\vec k), b^\dag(\vec{k}')]=(2\pi)^3\delta^{(3)}(\vec{k}-\vec{k}')$.

The classical equation of motion for $\zeta_{cl}$ follows from varying the quadratic $\pi_L$ action (\ref{longitudinal quadratic action}). The general solution to this equation is quite complicated, however there is a trick that makes its solution much easier. If we re-express the e.o.m. in terms of the other gauge invariant parameter $\mathcal{R}_{cl}$ (see eq.~\eqref{definition of adiabatic modes}) we have
\be
-3 c_L^2 \zeta_{cl} (\vec k, t) =\frac{1}{H}\dot{\mathcal{R}}_{cl}(\vec k, t)+(3+\eta(t)-2\epsilon(t))\mathcal{R}_{cl}(\vec k, t)\;. \label{equation of motion for scalar1}
\ee
which, together with the definition of $\mathcal{R}$ 
\be
\mathcal{R}_{cl}=\frac{1}{H \ep}\,\frac{\dot{\zeta}_{cl}+ H\ep\,\zeta_{cl}}{1+k^2/3a^2 H^2\ep} \;,\label{equation of motion for scalar2}
\ee
forms a system of two first order equations of two variables. Eliminating $\zeta_{cl}$ we can generate a second order equation for $\mathcal{R}_{cl}$ which takes the usual form similar to (\ref{tensoreom}). Written with respect to conformal time, and up to first order in slow roll, it is given by:
\be
\mathcal{R}''_{cl}+(2+\eta-2s) \, aH \;  \mathcal{R}'_{cl}+\left[c_L^2 k^2+(3\epsilon-6s+3 c_{L}^2 \epsilon) \, a^2H^2\right] \mathcal{R}_{cl}=0\;,
\ee
where prime denotes a derivative w.r.t. conformal time, and $s$ is the slow roll parameter defined by (\ref{s}). Once again, using the conformal time dependence of $aH$, $s$, $\eta$, $\epsilon$, and $c_L$ contained in Appendix \ref{time dep} this equation can be solved  in terms of Hankel functions. One finds that the general solution to first order in the slow roll parameters is given by
\begin{align}
\mathcal{R}_{cl}(\vec k, \tau)
&= \left(-\tau \right)^{-\alpha} \Big[\mathcal{C} H^{(1)}_{\nu_S} \big(- c_L(\tau) \, k \tau (1+s_c) \big) +\mathcal{D} H^{(2)}_{\nu_S} \big(- c_L(\tau) \, k \tau (1+s_c) \big) \Big] \label{R classical}
\end{align}
where $\alpha=-\frac{1}{2}(3+2\ep_c+\eta_c-2s_c)$ and $\nu_S=\frac{1}{2}(3+5s_c-2c_{L,c}^2\epsilon_c+\eta_c)$. 
Notice that for this to be a solution, it is important to keep into account---to first order---the time-dependence of $c_L$ 
in the argument of the Hankel functions.

Once again, in order to match the initial conditions we must canonically normalize $\pi_L$. A quick glance at (\ref{longitudinal quadratic action}) reveals that the correct canonically normalized field is 
\be
\pi_L^{\rm can.}(\vec k,\tau)=\sqrt{2}\,\pi_L \left[\frac{\mpl^2 a^2k^2   }{3\left(1+\frac{k^2}{3a^2 H^2 \epsilon}\right)}\right]^{1/2}  \stackrel{-k \tau \to \infty}{\longrightarrow} \sqrt{2\ep} \, \mpl H a^2 \, \pi_L \;.
\ee
With the usual normalization for the creation and annihilation operators we will recover the Minkowski vacuum for very early times by demanding that
\begin{align}
\lim_{\tau \rightarrow - \infty} \zeta_{cl}(\vec k,\tau)=-\frac{k \,\pi^{\rm can.}_{L,cl}}{3\sqrt{2\ep} \, \mpl H a^2}=-\sqrt{\frac{k}{4\, \ep c_L}}\frac{e^{-i (1+s_c)c_L(\tau) k \tau}}{3\,\mpl H a^2} \;. \label{Minkowski vacuum condition scalar1}
\end{align}
Or, equivalently that 
\be
\lim_{\tau \rightarrow - \infty} \mathcal{R}_{cl}(\vec k,\tau)=-\frac{a^2H^2}{k} \frac{d}{d\tau} \left( \frac{\pi^{\rm can.}_{L,cl}}{H}\right) =i\sqrt{\frac{c_L}{4\, \ep k}}\frac{e^{-i (1+s_c)c_L(\tau)k \tau}}{\mpl a}\; . \label{Minkowski vacuum condition scalar2}
\ee
Matching the general solution given by (\ref{R classical}) to the the initial condition (\ref{Minkowski vacuum condition scalar2}) will set $\mathcal{D}=0$ and
\be
\mathcal{C}=-i\sqrt{\frac{\pi}{8  \epsilon_{c}}}\frac{c_{L,c}H_c}{\mpl}(-\tau_c)^{s_c-\ep_c-\eta_c/2}(1+\sfrac12 {s_c}-\ep_c)e^{i(\eta_c+5s_c-2c_{L,c}^2\ep_c)\pi/4}+\mathcal{O}(\epsilon^{3/2}) \;.
\ee
%Here we have made sure to keep the dependence on $k$ to first order in slow roll as it will contribute to the tilt.
One can now use (\ref{equation of motion for scalar1}) to obtain the full expression for $\zeta_{cl}(\vec k, \tau)$, which is (as promised) a bit messy and not particularly instructive as for our computation we are only interested in $\zeta$'s late time limit. We will not bother to write it out here.  

Just like the tensor perturbations, neither $\mathcal{R}$ nor $\zeta$ is conserved outside the horizon, though their temporal dependence is 
mild, i.e., suppressed by slow-roll parameters:
\begin{align}
\lim_{-k\tau\to 0^+}\mathcal{R}_{cl}(\vec k,\tau)&=\left(\frac{\tau}{\tau_c}\right)^{\frac{4}{3}c_{T,c}^2\ep_c-2s_c}(-c_{L,c} k \tau_c)^{c_{L,c}^2\ep_c-5s_c/2-\eta_c/2}\left(-\frac{ H_c}{\sqrt{4\ep_c}\mpl c_{L,c}^{1/2}\,k^{3/2}}+\mathcal{O}(\ep^{1/2})\right)\label{asym scalar1}\\
\lim_{-k\tau\to 0^+}\zeta_{cl}(\vec k,\tau)&=\left(\frac{\tau}{\tau_c}\right)^{\frac{4}{3}c_{T,c}^2\ep_c}(- c_{L,c}k \tau_c)^{c_{L,c}^2\ep_c-5s_c/2-\eta_c/2}\left(\frac{H_c}{\sqrt{4\ep_c}\mpl c_{L,c}^{5/2}\,k^{3/2}}+\mathcal{O}(\ep^{1/2})\right)\label{asym scalar2} \;.
\end{align}
Notice that at this order in slow-roll, on large scales $\zeta$ and ${\cal R}$ are proportional to each other, with proportionality constant $c_L^2$, which is in agreement with (\ref{equation of motion for scalar1}):
\be \label{R and zeta}
{\cal   R} \simeq -c_L^2(\tau) \, \zeta \qquad (k\tau \to 0^-) \; .
\ee

Now finally, the two point function of  $\zeta$  for late times (when the modes are well outside the horizon) is given by
\begin{align}
&\big\langle  \zeta(\tau,\vec{k}_1) \zeta(\tau,\vec{k}_2) \big\rangle=(2\pi)^3 \delta^{3}(\vec k_1+ \vec k_2) \, \big| \zeta_{cl}(\tau,\vec k_1) \big|^2\\
&\stackrel{-k \tau \to 0^+}{\longrightarrow}(2\pi)^3 \delta^{3}(\vec k_1+ \vec k_2) \times 
\frac{H_c^2}{4 \epsilon_c c_{L,c}^5 \mpl^2} \, 
\frac{1}{k_1^3} \, 
\frac{(\tau/\tau_c)^{8c_{L,c}^2\ep_c/3}}{(-c_{L,c}k_1 \tau_c)^{5s_c-2c_{L,c}^2\epsilon_c+\eta_c}}  
\end{align}
where, as before, we have kept the dependence on $k$ and $\tau$ to first order in slow roll while the prefactor is expressed only to lowest order in slow roll.

Once again, since all the parameters are evaluated at the global time $\tau_c$ as opposed to the time of horizon crossing for each mode, we can simply read off the tilt  to first order in slow roll directly from the above expression. It is:
\be
\label{scalar tilt}
n_S-1 \simeq 2\epsilon_c c_{L,c}^2-5s_c-\eta_c \; .
\ee
Notice the overall $1/c_L^5$ factor in front of the spectrum. In a more standard single-field model, this would be replaced by $1/c_L$ (see e.g.~\cite{CCFKS}).
For small $c_L$, our extra powers of $c_L$ give us a very suppressed tensor-to-scalar ratio:
\be
\label{scalar to tensor ratio}
r \sim \epsilon \, c_L^5 \; .
\ee
It is crucial however to ascertain whether we should focus on the $\zeta\zeta$ spectrum or the ${\cal R} {\cal R}$ one. After reheating, when the universe is dominated by a hot fluid, they have to coincide, because of the usual reasons. But during inflation, because of \eqref{R and zeta}, they differ by a factor of $c_L^4$---precisely what suppresses our tensor-to-scalar ratio w.r.t~the standard case. In sect.~\ref{reheating} we argue that it is the $\zeta\zeta$ spectrum that is continuous at reheating.

%%%%%%%%%%%%%%%%%%%%%%%%%%%%%%%%%%
%%%%%%%%%%%%%%%%%%%%%%%%%%%%%%%%%%
\section{The three-point function}
\label{3ptfunction}

We now compute the $\langle \zeta \zeta \zeta \rangle$ three-point function. Like in single-field models with a small speed of sound, our three-point function will be enhanced by inverse powers of $c_L$ with respect to what one gets in standard slow-roll inflation, for essentially the same reason (see e.g.~\cite{CCFKS}). However in addition to this, we will find an extra $1/\epsilon$ enhancement, coming from the fact that the quadratic phonon action \eqref{kinetic} is suppressed by $\epsilon$, whereas the cubic interactions \eqref{cubic} are not.

%The three point function $\langle \zeta(\vec k_1,\tau) \zeta(\vec k_2,\tau) \zeta(\vec k_3,\tau)\rangle$ is of important experimental interest. Its presence (or lack thereof) and shape will be able to put constraints on the possible mechanisms for inflation \cite{}. For example, in the classic work of \cite{Maldacena} it was demonstrated that non-gaussianities in single field inflation models with a potential of the form $V(\phi)$ are suppressed by the slow roll parameter and therefore an observed non-guassianity greater than a specific value would rule out such a class of models. See sect. \ref{The size and shape of non-gaussianities} for more details.

In order to compute the correlation function at a specific time, we need to evolve it from a quantum state we know, that is the early-time flat-space vacuum. Expanding the usual time-evolution operator and working to lowest order in perturbation theory we have the standard result, which is given schematically by:
\be
\left < \zeta(\tau)^3\right>= -i \int_{-\infty}^\tau d \tau' \left<\Omega(-\infty) \left|\left[ \zeta(\tau)^3,H_{\rm int}(\tau')  \right ]\right|\Omega(-\infty)\right> \;.
\ee
For our purposes it is enough to calculate the three-point function to lowest order in slow-roll. As demonstrated in Appendix \ref{trilinear action}, at this order it is enough to work with the phonon cubic action \eqref{cubic}, which in our FRW curved background takes the form (neglecting boundary terms):
\be
\label{simplified L_3}
\mathcal{L}_{3} = \mpl^2 \, a(t)^3 H^2 \, \frac{F_Y}{F}\Big\{\sfrac{7}{81}(\pd \pi)^3-\sfrac{1}{9}\pd\pi \pd_j\pi^k\pd_k\pi^j-\sfrac{4}{9}\pd\pi \pd_j\pi^k\pd_j\pi^k +\sfrac{2}{3}\pd_j\pi^i \pd_j\pi^k\pd_k\pi^i\Big\}  \; .
\ee
Quite amazingly, this applies both in the decoupling limit ($k\gg aH \epsilon^{1/2}$) and in the opposite limit ($k\ll aH \epsilon^{1/2}$). And so, defining $\zeta_i \equiv \zeta(\tau, \vec k_i)$, we have
\begin{align}
\big< \zeta_1 \zeta_2 \zeta_3 \big> & = i \mpl^2 \frac{F_Y}{F}\frac{k_1 k_2 k_3}{27} \int_{\vec p_1, \vec p_2, \vec p_3} (2\pi)^3\delta^{3}(\vec{p}_1+\vec{p}_2+\vec{p}_3) \, Q(\vec p_1,\vec p_2,\vec p_3) \times \label{3ptintegral} \\
&\int_{-\infty}^\tau d\tau' \; a^4(\tau') H^2(\tau') \big<\big[\pi_L(\tau,\vec k_1)\pi_L(\tau,\vec k_2)\pi_L(\tau,\vec k_3),\pi_L(\tau',\vec p_1)\pi_L(\tau',\vec p_2)\pi_L(\tau',\vec p_3)\big]\big> \; , \nonumber
\end{align}
where
\begin{align}
Q(\vec{p}_1,\vec{p}_2,\vec{p}_3)\equiv & \; \frac{7}{81} \, p_1 p_2 p_3-\frac{5}{27} \, \Big(p_1\frac{(\vec{p}_2\cdot\vec{p}_3)^2}{p_2 p_3}+p_2\frac{(\vec{p}_1\cdot\vec{p}_3)^2}{p_1 p_3}+p_3\frac{(\vec{p}_1\cdot\vec{p}_2)^2}{p_1 p_2}\Big)\nonumber\\
\label{Q}
&+\frac{2}{3}\, \frac{(\vec{p}_1\cdot\vec{p}_2)(\vec{p}_2\cdot\vec{p}_3)(\vec{p}_3\cdot\vec{p}_1)}{p_1 p_2 p_3 } \; ,
\end{align}
which is totally symmetric under permutations of $\vec{p}_1,\vec{p}_2,\vec{p}_3$.

Writing $\pi_L$ in terms of creation and annihilation operators allows us to easily express the integral in terms of the classical solutions. To be precise,
\be
\big< \zeta_1 \zeta_2 \zeta_3 \big> = - (2\pi)^3\delta^{3}(\vec{k}_1+\vec{k}_2+\vec{k}_3) \times 6 \mpl^2 \frac{F_Y}{F}\frac{k_1 k_2 k_3}{27} \, Q(\vec k_1,\vec k_2,\vec k_3) \, I(\tau ;-\infty) \; ,
\ee
where the integral $I(\tau_1;\tau_2)$ is defined as 
\begin{align}
\label{Def of Integral}
I(\tau_1;\tau_2) & = J(\tau_1;\tau_2)+J^*(\tau_1;\tau_2)\\
J (\tau_1; \tau_2) &\equiv-i \, \pi_L^{cl}(\tau_1, \vec{k}_1)\pi_L^{cl}(\tau_1, \vec{k}_2)\pi_L^{cl}(\tau_1, \vec{k}_3) \int_{\tau_2}^{\tau_1} d \tau' \; \frac{(\tau'/\tau_c)^{-2\epsilon}}{H_c^{2}\tau'^4} \pi_L^{cl \, *}(\tau', \vec{k}_1) \pi_L^{cl \, *}(\tau', \vec{k}_2) \pi_L^{cl \, *}(\tau', \vec{k}_3) \nonumber
\end{align}
and we used that $Q$ is an even function of the momenta. Just as in the previous section, we have used the dependence of $H$ and $a$ on conformal time with reference to $\tau_c$ to lowest order in slow roll.

Utilizing the classical solution of $\mathcal{R}(\tau, \vec k)$ given by (\ref{R classical}) we can immediately recover $\pi_L^{cl}(\tau, \vec k)$.
This general solution, containing Hankel functions, and derivatives of Hankel functions is not particularly useful for attempting to perform the time-integral given by (\ref{Def of Integral}). We will first simplify $\pi_L^{cl}(\tau, \vec k)$, which will enable us to express the three-point function in an analytic form.

First, note that at high momenta, above the de-mixing scale the quadratic Lagrangian for $\pi_L$ takes the simpler form:
\be
S_{\rm demix}=\int_{\vec k} \int d\tau \; \big( \mpl^2 a^4 H^2\epsilon \big)\big[ \big|\pi_L' \big|^2 - c_L^2 k^2 \big| \pi_L\big|^2 \big] \; ,
\ee
which is---apart from the overall time-dependent pre-factor---the standard quadratic action for a scalar with generic propagation speed $c_L$.
Thus, one can solve the equations of motion generated by varying the above expression, fixing the exact form by requiring the flat-space vacuum in the infinite past (as we have done in the previous sections), without having to go through $\mathcal{R}$. Of course, the two methods are equivalent, but this route makes the method of expansion clear. As for momenta that are deep in the full-mixing limit, where $\left| c_L k \tau \right| \ll \epsilon^{1/2}$, $\mathcal{R}$ can be written in a simple form utilizing the asymptotic limit of the Hankel function (\ref{asym Hankel small x}),  which upon insertion into (\ref{equation of motion for scalar1}) yields a particularly simple $\pi_L=-3 \zeta / k$. 
In fact, even though in principle the two asymptotic expressions we get in this way should not share a common regime of applicability, in practice they are indistinguishable (at lowest order in slow-roll) in a wide range of momenta, $e^{- {\cal O}(1/\epsilon)} < \left| c_L k \tau \right| \ll 1$.
\footnote{This stems from the mildness of their time-dependence outside the horizon.}
Therefore, as a very good approximation to the full solution, we can use the two asymptotic solutions and `glue' them together anywhere inside this range. We find it  more convenient to glue them at $(c_L k \tau) \sim \epsilon$.
In summary, to lowest order in slow roll we can approximate the profile of $\pi_L^{cl}$ by:
\begin{subnumcases}{\pi_L^{cl}(\tau, \vec{k})\simeq}
\mathcal{B}_k\big(1+i c_L k \tau-\sfrac{1}{3}c_L^2k^2\tau^2\big)e^{-i c_L k \tau}\; ,& for $\vert c_L k \tau \vert \gtrsim \ep\qquad$ \label{pidemix}\\
\mathcal{B}_k \big(-c_{L,c} k \tau \big)^{c_{L,c}^2\ep_c+\epsilon_c} \big(-c_{L,c}k\tau_c \big)^{-5s_c/2-\eta_c/2-\ep_c} + {\cal O}(\ep) \; ,&for $\vert c_L k \tau \vert \lesssim \ep$\label{pimix}
\end{subnumcases}
where
\be
\mathcal{B}_k=- \frac32 \frac{H_c}{\mpl  c_{L,c}^{5/2} \ep_c^{1/2}} \frac{1}{k^{5/2}}\; .
\ee
%and where we have extended the solutions that should be valid only in the regions $\vert c_L k \tau \vert \gg \ep$ and $\vert c_L k \tau \vert \ll \ep$ all the way to $\vert c_L k \tau \vert \sim \ep$. 
Our strategy now is to break up the integral (\ref{Def of Integral}) into separate regions where one of the functional forms described by (\ref{pidemix}) and (\ref{pimix}) can used. The integral can then be done explicitly. 

%One may rightly worry that the intermediate region where $c_L k \tau \sim \ep$, which is not covered accurately by \eqref{pidemix} and \eqref{pimix}, will contribute $\mathcal{O}(1)$ corrections to our integral. Luckily, one can check this explicitly by numerically performing the integral over such intermediate regions using the full solution $\pi_L^{cl}(\tau, \vec k)$ (generated by the full solution of $\mathcal{R}_{cl}(\tau,\vec k)$) and comparing to the analytic result computed using the simplified solutions \eqref{pidemix} and \eqref{pimix}. In particular, we find for the simple quasi-equilateral case that the correct treatment of this region contributes negligibly. We will provide more specific details in the next section. 

\subsection{Analytic Calculation of Integral}
To illustrate the point and make the flavor of the analysis transparent, let's look at an almost equilateral configuration of momenta. That is, assume that 
\be
\vec k_1+\vec k_2+\vec k_3=0\;, \qquad  k_1 \sim k_2\sim  k_3 \sim k \; .
\ee
First, notice that given some reference time $\tau_*$, we can split the time-integral as
\begin{align}
\label{J identity}
J(\tau;-\infty) & =\frac{\pi_L^{cl}(\tau,\vec{k}_1)\pi_L^{cl}(\tau,\vec{k}_2)\pi_L^{cl}(\tau,\vec{k}_3)}{\pi_L^{cl}(\tau_*,\vec{k}_1)\pi_L^{cl}(\tau_*,\vec{k}_2)\pi_L^{cl}(\tau_*,\vec{k}_3)} \, J(\tau_*;-\infty)\\
& -i \, \pi_L^{cl}(\tau,\vec{k}_1)\pi_L^{cl}(\tau,\vec{k}_2)\pi_L^{cl}(\tau,\vec{k}_3)
\int^\tau_{\tau_*} d \tau' \; \frac{(\tau'/\tau_c)^{-2\epsilon_c}}{H_c^{2}\tau'^4}\pi_L^{cl \,*}(\tau',\vec{k}_1)\pi_L^{cl \, *}(\tau',\vec{k}_2) \pi_L^{cl \, *}(\tau',\vec{k}_3) \;. \nonumber
\end{align}
Then, choosing $\tau_*$ to be precisely the conformal time at which a mode of momentum $k$ transitions from \eqref{pidemix} to \eqref{pimix}, $-c_L k \tau_* \sim \epsilon$, we find that the real part of the second line vanishes at zeroth order in $\epsilon$, because all the $\pi_L$'s involved in the expression---inside and outside the integral---are real, and there is an overall $i$.
The remaining piece is all that will contribute to the integral. And so we can write
\begin{align}
J(\tau;- \infty)+J^*(\tau;- \infty)  = & \prod_{i=1}^3 \left|\mathcal{B}_{k_i}\right|^2 (-c_L k_i \tau)^{c_L^2\ep+\epsilon} (-c_{L}k_i\tau_c)^{-5s/2-\eta/2-\ep}\times \nonumber \\
& \int^{\tau_*}_{-\infty} d \tau' \; \frac{-i(\tau'/\tau_c)^{+\epsilon_c}}{H_c^{2}\tau'^4}\prod_{j=1}^3\big(1-i c_L k_j \tau'-\sfrac{1}{3}c_L^2k_j^2\tau'^2\big) \, e^{+i c_L k_j \tau'}+\text{c.c.}\nonumber \\
\label{J+Jstar}
= & -\frac1{27} \frac{ c_L^3}{H_c^2 } \, k_1 k_2 k_3 \, U(k_1,k_2,k_3)\left(\frac{\tau}{\tau_c}\right)^{4c_T^2\ep}\prod_{i=1}^{3}\vert\mathcal{B}_{k_i}\vert^2 (-c_{L}k_i\tau_c)^{c_L^2\ep-5s/2-\eta/2}\;, 
\end{align}
where the scale invariant function $U(k_1,k_2,k_3)$ is given by 
\begin{align}
\label{U}
U(k_1,k_2,k_3)&=\frac{2}{k_1k_2k_3(k_1+k_2+k_3)^3}\Big\{3 \big(k_1^6+k_2^6+k_3^6 \big) +20 \, k_1^2k_2^2k_3^2 \\
&+18 \big(k_1^4 k_2 k_3+ k_1 k_2^4 k_3+k_1 k_2 k_3^4 \big) +12 \big(k_1^3k_2^3+k_2^3k_3^3+k_3^3k_1^3 \big)\nonumber\\
&+9 \big(k_1^5 k_2+5 \text{ perms} \big)+ 12 \big(k_1^4 k_2^2+5\text{ perms} \big)+18 \big(k_1^3k_2^2k_3+5\text{ perms}\big) \Big\} \nonumber \; .
\end{align}
In order to ensure convergence of the integral and project onto the right vacuum, the integral is actually computed over a slightly tilted contour, that is $\tau' \rightarrow (1-i \varepsilon ) \tau '+\tau_*$, with $\varepsilon\to 0^+$, and the limits of integration are from $-\infty$ to $0$.  Additionally, in the last step, the fact that $1 \gg |c_L k \tau_* | \sim \epsilon$, and $ |\tau_*| > |\tau|$ has been used to collect only the leading order in slow roll contributions.

A more careful analysis of the same flavor applies to more general triangle shapes, see Appendix \ref{for general triangles}. It turns out that the above expression is valid provided that the the triangle formed by the various momenta is not too squeezed, that is, provided
\be
k_{\rm long} /k_{\rm short} > \sqrt{\epsilon} \; .
\ee 
%
%Once again, readers may be skeptical about our analysis. The valid regimes of the expressions for $\pi_L^{cl}$ (\ref{pidemix}) and (\ref{pimix}) should really be $\vert c_L k \tau \vert\gg\ep$ and $\vert c_L k \tau \vert\ll\ep$ respectively. Therefore, our analytic analysis above is justifiable only if the contribution to the integral (\ref{3ptintegral}) from the intermediate regime, $\vert c_L k \tau \vert\sim\ep$, is negligible. As discussed above, one can run some numerical consistency checks to verify that this is indeed the case, despite the fact that we are unable to give an analytical estimate of this intermediate regimes contribution to the integral.  In particular, for the simplest case where the momenta form a quasi-equilateral triangle, $k_1\sim k_2 \sim k_3$, we find $\int^{0.1\tau_*}_{10\tau_*}\sim 10^{-11} \int^{10\tau_*}_{-\infty}$. And so, while one should technically perform this check for all possible momentum configurations its success in this simple situation indicates that it is probably true for more general cases. This makes sense, as we can see from (\ref{J+Jstar}) that the main support over the integral is not near $\tau_*$ as it is not very sensitive to its exact value, i.~e.~ the integral would be the same to leading order in slow roll even if we stopped at, say, $5 \tau_*$.
%
%
And so finally, putting everything together, we can express the full three-point function as
\begin{align}
\label{three-point function}
\big< \zeta_1 \zeta_2 \zeta_3 \big>(\tau_e) & \simeq  (2 \pi)^3\delta^{3}(\vec{k}_1+\vec{k}_2+\vec{k}_3\big) \times \\ 
& \frac{3}{32}\frac{F_Y}{F}\frac{H_c^4}{\mpl^4} \frac{1}{\epsilon^3 c_L^{12}} \, \Big( \frac{\tau_e}{\tau_c}\Big)^{4 c_T^2 \epsilon}
  \, \frac{Q(\vec{k}_1,\vec{k}_2,\vec{k}_3) \, U(k_1,k_2,k_3)}{k_1^3\,k_2^3\,k_3^3} \; , \nonumber
\end{align}
where, we remind the reader, $Q(\vec{k}_1,\vec{k}_2,\vec{k}_3)$ is given by (\ref{Q}) and $U(k_1,k_2,k_3)$ is given by (\ref{U}).

We will remark that the mild time dependence, $(\tau_e/\tau_c)^{4c_T^2\ep}$, in the above expression can actually produce an order one correction to the overall magnitude of the three-point function.   Indeed, assuming that inflation lasts for $N_e\sim 60$ $e$-folds after the longest mode of today's relevance exits the horizon,  we can see immediately  that $(\tau_e/\tau_c)^{\mathcal{O}(\ep)}\sim e^{-60\times\mathcal{O}(\ep)}$, which, as promised, depending on how small $\epsilon$ is, can give an $\mathcal{O}(1)$ correction. 
On the other hand, the mild momentum dependence $(-c_L k_i \tau_c)^{c_L^2\ep-5s/2-\eta/2}$, appearing in (\ref{J+Jstar}), is equal to one up to $\mathcal{O}(\ep)$ corrections. We thus drop this piece from (\ref{three-point function}), in order to be consistent with the preceding computation of the integral.
Our result (\ref{three-point function}) should be understood as the leading order contribution in slow roll. 
%\an{Is the following true? You guys convinced me that you actually computed the ``tilt''.} 
%A word of clarification: the momentum dependence in (\ref{three-point function}) should be thought of as only supplying us with the necessary  overall scale. Since all momenta should be within a few orders of magnitude to $c_L^{-1}H_c$ there exact value is totally unimportant. That is, one should not think of their presence as some kind of ``tilt'' of the three-point function as we have not consistently kept all possible $k^\epsilon$ dependence that could arise from a higher order in slow roll treatment of the integral. 
%Therefore, to be consistent with the proceeding computation of the integral (\ref{J+Jstar}),  we will henceforth set the prefactor $(-c_L k_i \tau_c)^{c_L^2\ep-5s/2-\eta/2}=1$ in (\ref{three-point function}) while neglecting its $\mathcal{O}(\epsilon)$ correction. 

%%%%%%%%%%%%%%%%%%%%%%%%%%%%%%%%%%
%%%%%%%%%%%%%%%%%%%%%%%%%%%%%%%%%%
\section{The size and shape of non-gaussianities}
\label{The size and shape of non-gaussianities}

It is useful to rewrite the three-point function above as an overall amplitude  $f_{\rm NL}$ times  a {\em shape} that is a function of the momenta with order-one coefficients \cite{BCZ}. It is customary to do  so at the level of correlators of the Newtonian potential $\Phi$ during matter-domination, rather than of $\zeta$. The relation outside the horizon is
\be
\Phi = \sfrac35 \zeta \; .
\ee
Neglecting the tilt, one then writes the two- and three-point functions as  
\begin{align}
\langle \Phi(\vec k_1) \Phi(\vec k_2) \rangle & = (2\pi)^3 \delta^3(\vec k_1+ \vec k_2) \cdot \frac{\Delta_\Phi}{k_1^3} \\
\langle \Phi(\vec k_1) \Phi(\vec k_2) \Phi(\vec k_3) \rangle & = (2\pi)^3 \delta^3(\vec k_1+ \vec k_2+\vec k_3) \cdot f(k_1,k_2,k_3) \; ,
\end{align}
and normalizes $f$ on `equilateral' configurations with $k_1=k_2=k_3$ \cite{CNSTZ},
\be \label{def fNL}
 f(k,k,k) = f_{\rm NL} \cdot \frac{6 \Delta_\Phi^2}{k^6} \; .
\ee
This defines the parameter $f_{\rm NL}$ in a model-independent fashion, in terms of observable quantities only. In particular, the observed value for the power-spectrum normalization is $\Delta_\Phi \simeq 2 \times 10^{-8}$. Notice that, because of momentum conservation, the three momenta $\vec k_{1,2,3}$ close into a triangle. As a result, the function $f$ depends on the absolute values $k_{1,2,3}$ only, because a triangle is uniquely defined---up to overall rotations, which are a symmetry of $f$---by specifying its sides. Notice also that scale-invariance forces $f$ to have overall scaling dimension $k^{-6}$, and we have used this fact in \eqref{def fNL}. Notice finally that the standard convention would be to call $F$ the function that we call $f$. Unfortunately we have already been using $F$ for our Lagrangian, so will stick to $f$ for the function defined above. Hopefully this will not lead to confusion.

Applying these definitions to our case we find 
\begin{align}
\Delta _\Phi & = \frac{9}{100} \cdot \left(\frac{\tau_e}{\tau_c}\right)^{8 c_T^2\ep/3}\cdot \frac{H^2}{\mpl^2} \cdot \frac{1}{\epsilon \,  c_L^5} \\
f(k_1, k_2, k_3) & = \frac{5}{2} \cdot \frac{F_Y}{F}  \frac{1}{\epsilon \,  c_L^2} \cdot \left(\frac{\tau_e}{\tau_c}\right)^{-4 c_T^2\ep/3}\cdot \Delta^2_\Phi \cdot \frac{Q (k_1, k_2, k_3) \cdot U(k_1, k_2, k_3) }{k_1^3 k_2^3 k_3^3}  \label{f}\\
f_{\rm NL} & = -\frac{19415}{13122}\cdot\left(\frac{\tau_e}{\tau_c}\right)^{-4 c_T^2\ep/3} \cdot \frac{F_Y}{F}  \frac{1}{\epsilon \,  c_L^2} \simeq - \mathcal{O}(1)\cdot \frac{F_Y}{F}  \frac{1}{\epsilon \,  c_L^2}
\end{align}

The $f_{\rm NL}$ parameter gives us a measure of the absolute size of non-gaussianities. As we argued in sect.~\ref{inflation}, $F_Y$ is essentially a free parameter, which can be as large as $F$, in which case our $f_{\rm NL}$ is {\em huge}, of order $1/(\epsilon \, c_L^2)$. By comparison, single-field inflationary models with small sound speed---whose non-gaussianities are much larger than for standard slow-roll inflation---have, at the same value of the sound speed, an $f_{\rm NL}$ which is a factor of $\epsilon$ smaller than ours. 
Notice that in this case there is a potential tension for our model: the same combination $\epsilon \cdot c_L^2$ appears in the scalar tilt, eq.~\eqref{scalar tilt}. Of course one could have cancellations there, because of the other terms in the expression for the tilt. But assuming that these do not change the overall order of magnitude of the tilt, the tilt is small if non-gaussianities are large, and vice versa.  Eventually, one should observe either.
If on the other hand our $F_Y$ is of order $\epsilon F$, then this $1/\epsilon$ enhancement for non-gaussianities is gone, and our $f_{\rm NL}$ becomes of order $1/ c_L^2$, which is the same as for small sound-speed single-field models
\footnote{We remind the reader that cosmological perturbations can still be nearly gaussian, even for huge values of $f_{\rm NL}$, as long as the combination $f_{\rm NL} \sqrt{\langle \zeta^2 \rangle}$ is much smaller than one at the relevant scales. For us, in the most strongly coupled case ($F_Y \sim F$), such a combination is of order $H/\mpl \cdot (\epsilon c_L^3)^{-3/2}$, which is much smaller than one if and only if the weak-coupling condition \eqref{epsilon c^3} is obeyed. As usual, perturbations are  nearly gaussian if and only if they are weakly coupled at horizon crossing.}.

But the most interesting feature of our non-gaussian signal is probably its shape, that is, the dependence of $f$ on the shape of the triangle made up by the momenta $\vec k_{1,2,3}$. We plot it in fig.~\ref{shape}, following the standard conventions of \cite{BCZ}. In particular,  it is clear from the plot that our three-point function is peaked on `squeezed' triangles with $k_3 \ll k_{1,2}$, but its behavior for those configurations  depends strongly on the {\em angle} $\theta$ between $\vec k_3$ and the other momenta. Quantitatively, focusing on the $\frac{Q U}{k^3 k^3 k^3}$ structure in \eqref{f} and ignoring the prefactors from now on, we get
\be \label{squeezed}
f(k_1, k_2, k_3 \to 0) \propto -  \frac{40}{27} \frac{\big(1-3 \cos^2 \theta\big)}{k_1^3 k_3^3} \; .
\ee
where we used momentum-conservation to rewrite $k_2$ as $k_2 \simeq k_1 + k_3 \cos \theta$.
Such an angular dependence is not there in any of the standard inflationary models we are aware of: at least for single-field models, the consistency relations force the behavior of the three-point function in the squeezed limit to be angle-independent---see e.g.~\cite{Maldacena, CZ}. On the other hand, in our case the standard consistency relations  are maximally violated, both at the level of angular dependence---as we just mentioned---and at the level of the overall prefactor: usually the squeezed limit of the three-point function is suppressed by the scalar tilt, which is of order $\epsilon$; here instead, there is no suppression like that, and in fact, as we argued above the overall prefactor can be as big as one {\em over} the tilt. It is easy to see why the consistency relations do not hold in our model: the standard argument of \cite{Maldacena, CZ}---that a long-wavelength background $\zeta$ can be traded in for a rescaling of spatial coordinates---does not apply to our case, because in our model there is no gauge in which the curvature perturbation $\zeta$ appears as a $\zeta \cdot \delta_{ij}$ correction to the spatial metric.

The fact that our non-gaussianities feature a novel shape, means that the numerical analyses of CMB data that have been carried out so far are quite suboptimal for our case. Following \cite{BCZ}, this can be quantified by computing the overlap---or `cosine'---between our shape and those used in the analyses. These cosines give a measure of how much one could improve by performing a dedicating analysis, which would entail using directly our shape in the estimator for the $f_{\rm NL }$ parameter. Of all the standard shapes on the market, the {\em local} one is the only one that is peaked on squeezed triangles, with exactly the same scaling as ours, but with no dependence on the angle at which the squeezed limit is approached. As far as angular dependence goes, this corresponds to a monopole, whereas our squeezed limit \eqref{squeezed} corresponds to a quadrupole. As a result, the overlap between our shape and the local one vanishes in the squeezed limit, and overall it is very small
\footnote{According to the nomenclature of \cite{BCZ}, we are computing `3D' cosines, which are relevant for analyses of fully 3D data, like e.g.~those of large scale structure surveys. For analyses of CMB data, which are projected onto a 2D sphere, one should refer to 2D cosines, which are however much more cumbersome to compute. Even though they are not entirely precise in this case, 3D cosines are still a good  indicator of the overlaps  between different shapes.
We should also mention that for some of these cosines, the computations involve logarithmic divergences in the squeezed limit, which we cutoff at a value of $10^3$ for the maximum ratio between momenta in the data set.},
\be
\cos(f, f_{\rm local}) \simeq  2 \% \; .
\ee
Another popular shape is the {\em equilateral} one, which is peaked on configurations with $k_1= k_2 = k_3$. Our shape is fairly suppressed for those configurations, and this results in a smallish overlap:
\be
\cos(f, f_{\rm equil}) \simeq -39 \% \; .
\ee
Finally, given that our shape is fairly large for degenerate---or `flattened'---triangles with $k_1 = k_2 + k_3$, one might think that it has a large overlap with the so-called {\em orthogonal} shape. However, we find a small cosine with the orthogonal template of \cite{SSZ}:
\be
\cos(f, f_{\rm orthog}) \simeq -32 \% \; .
\ee
All this suggests that {\em (a)} at the level of the three-point function, our model is very distinguishable from more standard inflationary models, and that {\em (b)} to test our model, we should perform a dedicated CMB data analysis for our shape of non-gaussianities. 

%%%%%%%%%%%%%%%%%%%%%%%%%%%%%%%%%%
%%%%%%%%%%%%%%%%%%%%%%%%%%%%%%%%%%
\subsection{A factorizable template}

Numerical analyses of three-point correlation functions in the CMB drastically simplify if one can rewrite the theoretical momentum-space three-point function in a {\em factorizable} form, that is, as a sum of a few terms that are just products of powers of the triangle sides $k_1$, $k_2$,  $k_3$ \cite{KW}. If one is able to do so, the calculation time decreases by a factor of $N_{\rm pixel}$, which is a huge improvement.
Apart from isolated examples---most notably the so-called local form of non-gaussianities---factorization of a three-point function is typically possible only approximately.

We present here a factorizable form that is a very good approximation to our three-point function. As for other forms of non-gaussianities, finding an adequate factorizable approximation is a matter of trial and error. The zeroth order desiderata are: {\em (i)} it should have an overall $k^{-6}$ scale dependence; {\em (ii)} it should be totally symmetric in  $k_1$, $k_2$,  $k_3$;  {\em (iii)} it should have the correct squeezed limit. Notice that our squeezed limit, eq.~\eqref{squeezed}, {\em is} factorizable, since
\be
\cos \theta = \frac{\vec k_1 \cdot \vec k_2}{k_1 k_2} \; ,
\ee
and the dot product can always be rewritten, using $\vec k_1+ \vec k_2+ \vec k_3 = 0$, in terms of absolute values,
\be
2 \vec k_1 \cdot \vec k_2 = k_3^2-k_1^2-k_2^2 \; .
\ee
So our first (and final) guess, is to take the squeezed limit behavior and rewrite it in a totally symmetric form:
\be \label{Ffactor}
f_{\rm factor}(k_1, k_2, k_3) \equiv -   \frac{20}{27}  \bigg[\frac{k_1^2 k_3^2- 3 \big(\vec k_1 \cdot \vec k_3\big)^2}{k_1^5 k_3^5} + \mbox{2 permutations} \bigg] \; ,
\ee
where the factor of 2 difference with eq.~\eqref{squeezed} takes into account that for $k_3 \ll k_1, k_2$ we are getting the same contribution from the terms explicitly displayed and from one of the permutations.

\begin{figure}[t]
\begin{center}
\includegraphics[width=16cm]{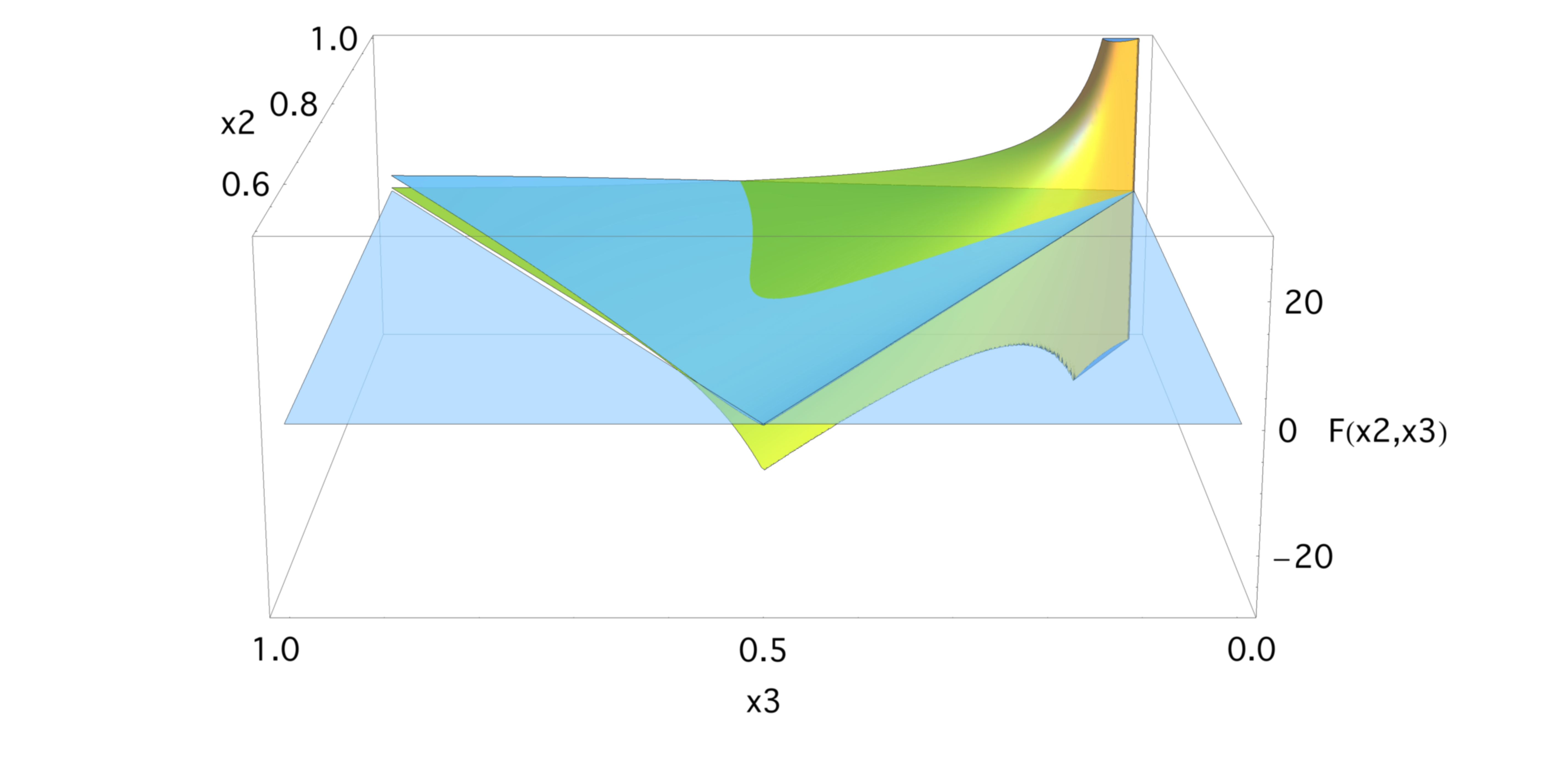}
\caption{\small \it Our factorizable approximation to the three-point function. The flat triangular surface is the difference with our exact three-point function. As clear from the picture, this is maximized for equilateral configurations ($x_2=x_3=1$), and is always quite small. \label{factorizable}}
\end{center}
\end{figure}

This ansatz obeys all the properties above, and performs surprising well in approximating our $f$ away from the squeezed limit. Quantitatively, for flattened configurations ($k_1 = k_2+k_3$)---where our three-point function has quite non-trivial features---the relative difference is very small, reaching a maximum of roughly $3 \%$ for `folded' triangles ($k_2=k_3= k_1/2$). For equilateral configurations the difference is more substantial, but still quite small in absolute terms, given that our signal is small there. We plot this factorizable form in fig.~\ref{factorizable}, alongside its
difference with the exact three-point function. The cosine between the two shapes is
\be
\cos (f, f_{\rm factor}) \simeq 97 \% \; .
\ee
If this level of precision is not enough, one can improve \eqref{Ffactor} by adding to it the right admixture of the equilateral factorizable form of \cite{CNSTZ}, to make up for the small difference in the equilateral limit:
\be
f_{\rm factor} \to f_{\rm factor}(k_1, k_2, k_3)  + f_{\rm equil}(k_1, k_2, k_3) \cdot \frac{f-f_{\rm factor}}{f_{\rm equil}}\bigg|_{k_1=k_2=k_3} \; .
\ee 
This  way the cosine with $f$  rises to roughly $99.5 \%$.

%%%%%%%%%%%%%%%%%%%%%%%%%%%%%%%%%%
%%%%%%%%%%%%%%%%%%%%%%%%%%%%%%%%%%
\section{Why Is $\zeta$ Not Conserved?}
\label{Why is zeta not conserved?}

We saw in sect.~\ref{2ptfunction} that already at linear level, during our solid inflation phase neither $\zeta$ nor ${\cal R}$ is conserved on large scales. One might be tempted to attribute this to the presence of isocurvature modes in addition to adiabatic ones. However, in our model there is only one scalar perturbation---parameterized by $\pi_L$ in the gauge we have been using---and usually isocurvature modes are a luxury that only multi-field models can afford. To sharpen the paradox,  our $\zeta$ and ${\cal R}$ do not coincide on large scales---see eq.~\eqref{R and zeta}---whereas usually they do, even for fluctuations that are not purely non-adiabatic, that is, even when they are not conserved.
In fact, Weinberg proved a no-go theorem stating that {\em all} FRW cosmological models---inflationary or not---feature two adiabatic modes of fluctuation, one of which has constant and identical $\zeta$ and ${\cal R}$ on large scales, while the other has $\zeta = {\cal R} = 0$ \cite{Weinberg_Adiabatic_Modes,Weinberg}. 
This theorem is manifestly violated by our model.
But, in Weinberg's own words, ``no-go theorems have a way of relying on apparently technical assumptions that later turn out to have exceptions of great physical interest" \cite{WeinbergCC}. We do not know whether our model will ultimately turn out to be of great physical interest, but it certainly offers an exception to an apparently technical assumption of ref.~\cite{Weinberg_Adiabatic_Modes}. Before showing this, let us explain in physical terms why our solid system cannot sustain adiabatic modes. 

By definition, an adiabatic mode is a perturbation that for very long wavelengths becomes locally unobservable, being indistinguishable from a slight shift in time of the background solution. An ordinary fluid offers a perfect example of this. Consider a  long-wavelength sound wave in a fluid, for the moment in the absence of gravity. For an observer making measurements on  scales much shorter than the wavelength, and working in the local rest frame---which is slightly different from the background one---, the only observables are the density and the pressure: neglecting gradients, a fluid is isotropic in its rest frame, and its stress-energy tensor is characterized by $\rho$ and $p$ only, which are related by the equation of state. Then, the only physical effect that is measurable on scales much shorter than the wavelength is the local compression (or dilation) the sound wave induces.
When we include gravity into the picture, essentially the same considerations apply for the perturbation, but now the time-evolution of the {\em unperturbed} FRW background already probes  all possible compression levels for the fluid (within some range), that is, all possible values of $\rho$ and $p$ compatible with the equation of state. As a consequence, for wavelengths much longer than the Hubble scale, within any given Hubble patch a sound wave will be indistinguishable from a time-shift of the background solution, that is, it will become physically unobservable.
Different Hubble patches will then evolve as separate identical FRW universes, and it can be shown that this translates into a conservation law for ${\cal R}$ \cite{BST}.

For a solid, the situation is very different, already in the absence of gravity. A longitudinal phonon---which is the only scalar fluctuation at our disposal---does not correspond to a purely compressional deformation of the medium. Even for wavelengths that are much longer than the observation scale, the anisotropic stress and the compression associated with the phonon are of the same order of magnitude. This is evident from the form of the stress-energy tensor \eqref{Stress Tensor}, which expanded to first order in $\vec \pi$ yields schematically 
\be \label{Tij solid}
\delta T_{ij} \sim (\vec \nabla \cdot \vec \pi) \, \delta_{ij} +  (\di_i \pi_j + \di_j \pi_i) \; ,
\ee
with similar coefficients in front of the two tensor structures---related to suitable derivatives of $F$ w.r.t.~$X,Y, Z$---whose precise values will not concern us here. For a longitudinal phonon of momentum $\vec q$, the anisotropic stress is proportional to $\hat q_i \hat q_j$, and is of the same order of magnitude as the change in pressure. A local observer can detect this anisotropy if he or she can detect the change in pressure. In other words, unlike a fluid,  a solid with small longitudinal deformations is not locally isotropic. Once we include gravity, these scalar fluctuations will be locally distinguishable from the background solution, even for super-horizon wavelengths, since the background {\em is} isotropic. One could probably apply a `separate universe'-like argument that includes anisotropic homogeneous backgrounds, to get information about the time-evolution of these modes, even at non-linear order
\footnote{We thank Matias Zaldarriaga for this remark.}.
We leave this for future work. For the moment, we just notice that our scalar modes do not conform to the standard characterization of adiabatic perturbations.

We can now go back to the no-go theorem of ref.~\cite{Weinberg_Adiabatic_Modes}, and see which assumptions we are violating.
The theorem is based on the following ingenious idea. Since an adiabatic mode is, by definition, unobservable once the wavelength is very long,
at zero momentum it should reduce to a gauge transformation of the FRW background. Newtonian gauge is a complete gauge-fixing at finite momentum, but it has a residual gauge freedom at zero momentum. By exploiting this gauge freedom one can construct zero-momentum solutions of the linearized perturbation equations. Most of these are pure-gauge, unphysical solutions. To be physical, they have to be the zero-momentum limit of finite momentum solutions, which are physical because there is no residual gauge-freedom at finite momentum.
For this to be the case, one needs the zero-momentum solution to obey the finite-momentum version of the $(ij)$ and $(0i)$ linearized Einstein equations,
\be \label{stronger einstein}
\Phi= \Psi - (8\pi G) \, \delta \sigma \; \qquad \dot H \, \delta u = \dot \Psi + H \Phi \; ,
\ee
which singles out only two independent modes among the zero-momentum solutions. Here $\delta \sigma$ and $\delta u$ are the scalar anisotropic stress and the velocity potential of eqs.~\eqref{Ti0}, \eqref{Tij},
\be \label{anisotropic}
\delta T_{ij}\supset a^2(t) \,  \di_i\di_j \delta \sigma \; , \qquad \delta T_{0i} \supset - (\bar \rho + \bar p) \, \di_i \delta u \; .
\ee 
One has to impose further that all equations of motion---for gravity and for the matter fields---are regular for $\vec q \to 0$, including eq.~\eqref{stronger einstein}. More precisely, if one rewrites all linearized equations of motion in first-order
form,
\be \label{system}
\dot y_a(\vec q, t) + C_{ab}(\vec q, t) \cdot y_b(\vec q, t) = 0 \; ,
\ee
where the $y_a$'s include the fields and their velocities,  with constraints for the initial conditions of the form
\be
c_b(\vec q \, ) \cdot y_b(\vec q, t_0) = 0 \; ,
\ee
then one has to demand that all $C_{ab}$ and $c_b$ coefficients be regular for $\vec q \to 0$. This is the only technical assumption of the theorem
\footnote{There is also an implicit assumption---used to write down the zero-momentum pure gauge solutions---that all background matter fields only depend on time, which is not obeyed in our case. However, this is easily fixed by performing the correct gauge transformation, which, following the notation of ref.~\cite{Weinberg_Adiabatic_Modes}, in our case yields the pure gauge solution
\be
\Psi=H \varepsilon - \lambda \; ,\qquad \Phi = - \dot \varepsilon \; , \qquad \vec \pi = \lambda \, \vec x \; , \qquad \delta \rho= - \dot{ \bar\rho} \, \varepsilon \; , \qquad \delta p= - \dot{ \bar p} \, \varepsilon \; .
\ee
}.
If it is obeyed, then the two zero-momentum gauge modes can be promoted to physical, finite-momentum solutions, which are adiabatic by construction, and one of which turns out to have constant $\zeta $ and ${\cal R}$.

From our physical argument above, we see that the culprit for us is the anisotropic stress, which does not become negligible at long wavelengths. Indeed, comparing \eqref{anisotropic} with \eqref{Tij solid} we get schematically
\be
\delta \sigma \sim \frac{\vec q}{q^2} \cdot \vec \pi
\ee
(we are neglecting factors of $a(t)$, and corrections to \eqref{Tij solid} involving the metric, which does not change our conclusion.)
Once plugged into eq.~\eqref{stronger einstein}, this gives us an equation of motion that is not regular for $\vec q \to 0$, thus violating Weinberg's technical assumption.  We get a similar singularity in the second equation of \eqref{stronger einstein}, since from $T_{0i} \sim \dot \pi^i$ we get a velocity potential
\be
\delta u \sim \frac{\vec q}{q^2} \cdot \dot {\vec \pi}
\ee
Notice that we cannot reabsorb the annoying $q^{-2}$ factors into a new $\vec \pi$ field, thus making $\delta \sigma$ and $\delta u$ regular for $\vec q \to 0$, because one of the equations \eqref{system} is the equation of motion for $\vec \pi$ itself, which is local in real space, thus analytic in $\vec q$ in Fourier space. If we were to divide it by $q^2$, to write an evolution equation for the new $\vec \pi/q^2$ field, we would introduce singularities there.

\section{Matching of  correlation functions at reheating, and post-inflationary evolution}\label{reheating}
Since our scalar perturbations are not adiabatic, our predictions for post-inflationary correlation functions on large scales can in principle be affected by local physical processes happening at reheating. It would be interesting to investigate how much model-independent information about time-evolution across the reheating phase can be obtained by applying an {\em anisotropic} parallel-universe argument, as mentioned in the last section. For the time being, we adopt what appears to be a reasonable model for reheating. As motivated in sect.~\ref{clock}, we postulate that inflation ends when $B\equiv \det B^{IJ}$ reaches some critical value $B_e$, after which the matter content of the universe turns into a perfect fluid, which can be described by a low-energy effective Lagrangian involving the same set of dynamical d.o.f.'s as our solid, but with  different (and more restrictive) symmetry requirements.  Furthermore, we assume that the transition (a.k.a. reheating) from solid to fluid occurs ``smoothly'' (in a sense that will be made explicit) and instantaneously (the transition time being much shorter than the Hubble time).  

Since it is the scalar quantity $B$ that plays the role of the ``clock'' in our model, it is easiest to work in unitary gauge (UG), in which constant time slices correspond to surfaces of uniform $B$ (the properties of this gauge are worked out in Appendix \ref{gauge app}). Explicitly, our model for reheating can be captured by the following statements:
\begin{itemize}
\item Inflation ends at $t_e$, with $a(t_e)^{-6}=B_e$, where $t$ denotes the time in UG.
\item The matter content in the post inflationary era takes the form of a perfect fluid, which can be described by the effective action \cite{DGNR, ENRW}
\be
S_{\rm fluid}=\int d^4 x \sqrt{-g}\, \tilde{F}(B)
\ee
where $B=\det g^{IJ}$ in UG. The change from one equation of state to the other is effectively instantaneous. For an ultra-relativistic fluid with $p = \frac13 \rho$, one has \cite{ENRW} 
\be
\tilde F(B) \propto B ^{2/3} \; .
\ee

\item Energy transfer from the solid phase to the fluid phase during this short reheating period is efficient, and the normalization of $\tilde{F}$ is restricted in such a way that energy conservation is respected, i.e. 
\be
\rho_{\rm solid}=-F(X,Y,Z)\big\vert_{t_e}=-\tilde{F}\left(a(t_e)^{-6}\right)=\rho_{\rm fluid} \; .
\ee
However, generally $p_{\rm solid} \ne p_{\rm fluid}$, since the equation of state has been changed.  Consequently, even though the Hubble parameter $H$ remains continuous, $\dot{H}$ does not.
\item Smoothness: the dynamical d.o.f.'s in unitary gauge---$A$ (or $\chi$), $C_i$, $D_{ij}$---as well as their first   derivatives are continuous across the $t=t_e$ surface
\footnote{See Appendix \ref{gauge app} for the definition of these fields in terms of fluctuations of the metric.}.  
The second derivatives will exhibit discontinuities since the equations of motions are altered due to the instantaneous change in the equation of state. 
\end{itemize}
An immediate consequence following the smoothness requirement is that $\zeta$  transits continuously from solid phase to post-inflationary fluid phase, while $\mathcal{R}$ does discontinuously. Indeed notice that in UG, $\zeta$ and $\mathcal{R}$ are given by
\be
\zeta=\frac{A}{2}, \quad \mathcal{R}=-\frac{H}{2\dot{H}}\frac{\dot{A}-\dot{H} A/H}{1-k^2/3a^2\dot{H}}
\ee
thus the discontinuity of $\mathcal{R}$ stems from that of $\dot H$.
%$\dot{H}=\ddot{a}/a^2-\dot{a}^2/a^2$.

Given that the transition from the solid phase during inflation to the perfect fluid phase during the post-inflationary era occurs effectively instantaneously, we can compute various correlators (of $A$, $C_i$ and $D_{ij}$) at $t_e$ and use them as the {initial conditions} of the post-inflationary evolution. 
There are two subtleties: 
\begin{enumerate}
\item How do we relate the correlators of quantities in UG to those in SFSG, which have been computed in Section \ref{2ptfunction} and Section \ref{3ptfunction}.
\item Given that the super-horizon modes of $A$ and $D_{ij}$ are not adiabatic during inflation, they start the post-inflationary evolution with a non-vanishing first derivative in time.  Although eventually these super-horizon modes become constant (time-independent) before reentering the horizon, a natural question to ask is how much the eventual constants could differ from the initial conditions these modes start with.
\end{enumerate}

In order to address the first issue, let's compute the scalar two- and three-point correlators: $\langle A_1(\bar{t}_e) \, A_2(\bar{t}_e)\rangle$ and $\langle A_1 (\bar{t}_e) \, A_2(\bar{t}_e) \, A_3 (\bar{t}_e)\rangle$, where the ``$~\bar{}~$'' is to remind ourselves that we are using time in UG, and $A_i(\bar t \,)$ is shorthand for $A(\vec k_i, \, \bar t \,)$.  
Using the transformation rule from SFSG to UG, we have that 
\begin{equation*}
\bar{t}=t-\frac{1}{3H}\pd_i\pi^i(x)+\mathcal{O}(\pi^2)\;,
\end{equation*}
and we can write
\begin{align*}
&A(x)=A^{(1)}+A^{(2)}+\dots,\\
&\text{where } \quad A^{(1)}(x)=\sfrac{2}{3}\pd_i\pi^i(x)=2\zeta,\qquad A^{(2)}\sim \pd\pi\pd\pi\;, \qquad \text{ etc.}
\end{align*}
It follows immediately that, schematically,
\begin{align}
\langle A^2\rangle=  \langle A^{(1)} A^{(1)}\rangle+2\langle A^{(1)} A^{(2)}\rangle+...\sim 4\langle\zeta^2\rangle+\langle(\pd\pi)^3\rangle+... \; .
\end{align}
The second term on the right can be neglected, since it is of higher order in the perturbative expansion. Likewise, we have 
\be
\langle A_1(\bar{t}_e) A_2(\bar{t}_e)\rangle \simeq\langle A^{(1)}_1(\bar{t}_e) A^{(1)}_2(\bar{t}_e)\rangle=4\langle \zeta_1(\bar{t}_e)\zeta_2 (\bar{t}_e)\rangle  \simeq 4\langle\zeta_1 (t) \zeta_2 (t)\rangle \label{2ptA}
\ee
where the last (approximate) equality is justified as long as the perturbative expansion (in fields) holds, since the difference between $t$ and $\bar t$ is of first order in the fields.  

We can do the same for the 3-pt correlators: 
\be
\langle A^3\rangle\simeq \langle A^{(1)} A^{(1)}A^{(1)}\rangle+3\langle A^{(1)} A^{(1)}A^{(2)}\rangle\sim 8\langle\zeta^3\rangle+\langle\zeta^2\pd\pi\pd\pi\rangle \; .
\ee
Notice that $\langle\zeta^2\pd\pi\pd\pi\rangle\sim \langle\pd\pi\pd\pi\rangle^2\sim O(\ep^{-2})$, while $\langle \zeta^3 \rangle \sim O(\ep^{-3})$, thus if we restrict ourselves to the leading order in slow-roll, this term can be safely neglected. It follows that
\begin{align} \label{3ptA}
\langle A_1(\bar{t}_e)A_2(\bar{t}_e)A_3(\bar{t}_e)\rangle & \simeq 8\langle\zeta_1(\bar{t}_e)\zeta_2(\bar{t}_e)\zeta_3(\bar{t}_e)\rangle \\ 
&\simeq 8\langle\zeta_1({t})\zeta_2({t})\zeta_3({t})\rangle +\frac{8 k_1}{3 H}\langle\dot{\zeta}_1(t) \, \pi_{L,1}(t)\zeta_2(t)\zeta_3(t)\rangle+{\rm perms.}\nonumber\\
&\simeq8\langle\zeta_1({t})\zeta_2({t})\zeta_3({t})\rangle +\ep \cdot \frac{32 c_T^2}{3}\langle {\zeta}_1(t) \, \zeta_{1}(t)\zeta_2(t)\zeta_3(t)\rangle+{\rm perms.}\nonumber \; ,
\end{align}
where we have used that in the long wavelength limit $\dot{\zeta} \simeq \zeta'/a\simeq \frac{4}{3}c_T^2\ep H \, \zeta$.
In the last line, the second term and its permutations are negligible at the leading order in slow roll, since $\ep\langle\zeta^4\rangle\sim \ep \langle \zeta^2 \rangle ^2 \sim O(\ep^{-1})$ while $\langle\zeta^3\rangle\sim O(\ep^{-3})$. 

Hence, as long as we focus only on the leading contribution in slow roll, the first issue mentioned above can be easily resolved: the 2-pt and 3-pt correlators of scalar perturbations in UG are related to those of (2 times) $\zeta$ in SFSG. Not surprisingly, similar relations for tensor perturbations hold if we apply the same logic.

As for the second issue. It can be shown \cite{LMS, LF} that during the post inflationary era, when the matter content of the universe is in the form of a perfect fluid, the scalar perturbation $A$ is adiabatic in the long wavelength limit, i.e. it is a constant at nonlinear level as long as it stays outside the horizon.
However, unlike other inflationary models where there exists a conserved scalar mode in the long wavelength limit {\em during} inflation, the scalar perturbation $A$ in our model evolves slowly outside the horizon, in the sense that $A\simeq A^{(1)}=2\zeta\propto (-\tau)^{\frac43 c_T^2\ep}$. Therefore, after the rapid transition from solid phase to perfect fluid phase, rather than staying at its initial value, $A(t_e)$, it approaches its eventual constant value. However the relative difference between the two is only of order $\epsilon$: the slow time-dependence of $A$ during inflation means that right after reheating the initial condition for the velocity is roughly $\dot A(t_e) \sim \epsilon  \cdot H A(t_e)$. Since then, $\dot A$ decreases like $1/a^3$, thus making $A(t)$ approach its asymptotic value in a few Hubble times, during which $A(t)$ moves by $\sim \epsilon A(t_e)$.
At the leading order in slow-roll, we can neglect this difference. Notice that this effect cannot change the tilts that we have computed: all modes of interest are outside the horizon during reheating and during the phase when $A$ relaxes to its asymptotic value. As a result, this small correction of order $\ep$ to the value of $A$ is the same for all modes, i.e., independent of  $k$.
%
%Let us estimate how far away from the initial value the eventual constant value end up. Notice that as
%\be
%A(\vec{k},t)\simeq A^{(1)}(\vec{k},t)=\int ^t _{t_e} d\,t'\frac{\dot{A}^{(1)}(\vec{k},t_e)a(t_e)^3}{a(t')^3}+A^{(1)}(\vec{k},t_e), \quad k\ll a H \label{postevoA}
%\ee
%and as the scale factor $a(t)$ grows (as a power law for instance, if $w\equiv p/\rho=$ const.), $A^{(1)}$ indeed approaches its asymptotic value rapidly.  Furthermore, given
%\be
%\frac{\dot{A}^{(1)}(\vec{k},t_e)}{A^{(1)}(\vec{k},t_e)}\bigg\vert_{UG}=\frac{2k/3\, \dot{\pi}_{cl}(\vec{k},t_e)}{2k/3\,\pi_{cl}(\vec{k},t_e)}\bigg\vert_{SFSG}\simeq -\frac{4c_t^2}{3}H(t_e)\ep
%\ee
%we would expect that
%\be
%\bigg\vert\frac{A^{\text{asym}}(\vec{k})-A^{(1)}(\vec{k},t_e)}{A^{(1)}(\vec{k},t_e)}\bigg\vert\sim \mathcal{O}(\ep) \;.
%\ee
%Therefore, at the leading order in slow roll, $\langle A^{\text{asym}}(\vec{k}_1)\dots\rangle\sim \langle A(\vec{k}_1,\bar{t}_e)\dots\rangle\sim 2^n\langle \zeta(\vec{k}_1,t)\dots\rangle$. In this way, we can estimate the quantities that could be potentially compared with the CMB data --- the correlators of $A^{\text{asym}}$ --- via those correlators computed in Section \ref{2ptfunction} and Section \ref{3ptfunction}.
By applying the same logic, we can reach the same conclusion for the transverse traceless tensor perturbation $D_{ij}$, which is not conserved in the long wavelength limit during inflation, but approaches an asymptotic value in the post-inflationary fluid phase in a similar manner as its scalar counterpart. 

%Before ending this section, we would like to remark on an interesting parametric regime where $c_L/c_t\to 0$, as %a consequence of which the transverse vector modes are suppressed and barely produced during inflation %\footnote{Equivalently,  we can start our model construction demanding the absence of the transverse vector %modes by postulating that $\phi^I=x^I+\pd^I \pi$.}. This can be seen by noting that $\vert %\pi_{T,cl}^i(\tau,t_e)/\pi_L^{cl}
%(\vec{k},t_e) \vert \simeq \left(c_L/c_t\right)^{5/2}$.  Thus after reheating, there will be no vortex modes present in the perfect fluid phase; according to the arguments of \cite{}, such fluid can be %described by the effective superfluid action
%\be
%S_{sf}=\int {\rm d}^4x \,P(\pd_\mu\psi\pd_\nu\psi g^{\mu\nu})
%\ee
%where the ground state of $\langle\psi\rangle=t$ spontaneously breaks the temporal differomophism. 

In conclusion, at the order we are working, we can take the correlation functions for $\zeta$ and $\gamma$ that we have computed during inflation in SFSG, evaluate them right before reheating, and obtain in this way good approximations to the corresponding correlation functions in UG in the post-inflationary phase. In particular, even though our scalar perturbations are not adiabatic during inflation, at reheating they get converted to adiabatic ones, with the same asymptotic constant value of $\zeta$ (up to $O(\ep)$ corrections) as they had at reheating.

%%%%%%%%%%%%%%%%%%%%%%%%%%%%%%%%%%
%%%%%%%%%%%%%%%%%%%%%%%%%%%%%%%%%%
\section{Summary and outlook}\label{conclusions}
\label{Summary}

Our model differs drastically from more standard ones in its symmetry breaking pattern. In particular, time-translations are not broken:
there are physical ``clocks''---i.e., time-dependent gauge-invariant observables---, but they inherit their time-evolution  from the metric, not from the matter fields. As a result, the systematics of the  EFT for the associated Goldstone excitations is completely different than the standard effective field theory of inflation. This has far reaching implications, some of which are directly observable.

The observational predictions of our model can be summarized as follows:
{\em (i)}
A nearly scale-invariant spectrum of adiabatic scalar perturbations, in agreement with observations.
{\em (ii)}
A nearly scale-invariant spectrum of tensor modes, with a slight {\em blue} tilt; the tensor-to-scalar ratio $r \sim \epsilon c_L^5$ ranges from somewhat smaller than in standard slow-roll inflation, for ultra-relativistic longitudinal phonons with $c_L^2 \simeq 1/3$, to tiny, if they are non-relativistic.
{\em (iii)} A  scalar three-point function with a novel shape---peaked in the squeezed limit, with non-trivial angular dependence on how the limit is approached---and a potentially very large amplitude, as big as $f_{\rm NL} \sim \frac{1}{\ep} \frac{1}{c_L^2}$.

A fourth prediction, whose full analysis we leave for future work, is the presence of vector modes. Unlike  usual inflationary systems, our solid features transverse phonons, which get excited during inflation. In Appendix \ref{vectors} we show that the vector-to-scalar ratio at the end of inflation scales as $(c_L^2/c_T^2)^{5/2}$, which, depending on the value of $c_L$, ranges from negligible to roughly $6 \%$. After reheating, the universe is dominated by a perfect fluid, and these vector modes should decay in the usual fashion, thus leaving no detectable imprint on the CMB. However, they interact at non-linear order with scalar and tensor perturbations during inflation and at reheating, thus affecting in principle their (higher point) correlation functions. 

Also for future work we leave a more thorough understanding of the time-evolution of our perturbations. One should be able to get model-independent information by running a separate universe-type argument involving anisotropic background solutions. In particular, this approach might elucidate whether and how super-horizon perturbations get affected by local processes at reheating. 
We showed in sect.~\ref{reheating} that a quick phase-transition triggered by $\det B^{IJ}$  conserves $\zeta$ and the tensor modes.
Moreover, we showed that for the two- and three-point function, there is no difference between using unitary gauge time and SFSG time, which implies that we would get the same post-inflationary correlation functions if reheating were in fact triggered   by another ``clock'', say $[B]$. This gives us confidence that, at least for these correlation functions, our predictions are robust. However it leaves open the possibility that we would get different predictions if the reheating phase lasted longer, for an Hubble time or more.

There is also a number of generalizations of our model that we feel deserve being studied.
The first, would be to promote our solid to a `super-solid'. In physical terms, a super-solid is a solid harboring a superfluid \cite{AL}. 
In our field-theory, symmetry-breaking terms, it is a system of four derivatively coupled scalars, $\phi^0$ and  $\phi^I$, with a shift-symmetry on $\phi^0$, our solid symmetries on the $\phi^I$'s, and a state that spontaneously breaks {\em all} space-time translations as follows \cite{Son}:
\be
\langle\phi^0\rangle = \mu \, t \; , \qquad \langle \phi^I \rangle = \alpha \, x^I \; .
\ee
This system  then combines the symmetry-breaking pattern of standard inflationary models with ours, and provides the minimal ingredients---the analog of `single-field'---for studying the consequences of breaking all space-time translations during inflation, in a way that is consistent with a residual physical homogeneity and isotropy. Cosmological systems of this sort have been considered briefly in \cite{DGNR}.
Notice that such a model will have {\em two} scalar perturbations---roughly speaking, excitations of $\phi^0$ and longitudinal excitations of $\phi^I$, although in general they will be mixed, like it happens for instance in finite-temperature superfluids \cite{Nicolis}, where the two resulting modes go under the names of first and second sound.
Notice also that, for inflationary purposes, one does not need to assume that the shift-symmetry on $\phi^0$ be exact. One can assume an approximate shift-symmetry, with symmetry-breaking couplings suppressed by slow-roll parameters---in which case, to be consistent with the literature, one should refrain from calling such a system a super-solid.

Another generalization that one should consider, is demoting our solid from isotropic to {\em crystalline}. We say ``demoting'', because it would entail lowering the degree of internal symmetries acting on our $\phi^I$ fields, from the full $SO(3)$ group to one of its discrete subgroups, e.g.~the cubic symmetry group. Then, in the background configuration with $\langle \phi^I \rangle = \alpha x^I$, only that particular discrete subgroup will be preserved (now as a linear combination of internal and spatial rotations), and one could have interesting, observable deviations from isotropy. One of course needs to make sure that these deviations are absent from the background evolution and from the spectrum of perturbations---which we know to be isotropic with very good accuracy---but this might be automatic for certain subgroups of $SO(3)$. As an example, consider the background evolution. The potential anisotropy in it is determined by the tensor structure of $T_{ij}$. Let's assume for definiteness that our discrete subgroup is the cubic group---by which we also mean the individual inversions along the sides---and let's align the sides of the cube with $\hat x$, $\hat y$, and $\hat z$. Then, $T_{ij}$ has to be invariant under permutations of $\hat x$, $\hat y$, and $\hat z$, and under the individual parities $\hat x \to - \hat x$, etc. The only two-index tensor with these properties is
\be
\hat x^i \hat x^j + \hat y^i \hat y^j +\hat z^i \hat z^j \; ,
\ee
which is nothing but $\delta^{ij}$. That is, at the two-index level, cubic symmetry accidentally implies full $SO(3)$ invariance. For the spectrum of perturbations to be accidentally isotropic in the same fashion, one needs the phonons' quadratic action to be accidentally isotropic. This involves now four-index background tensors, contracted with the phonons' derivatives. So, the general mathematical question is, what are the  discrete subgroups of $SO(3)$ whose invariant two-index and four-index tensors are all isotropic? If one finds one or more subgroups with this property, and if their {\em six}-index invariant tensors are {\em not} all isotropic, one has a model of inflation, that, because of symmetry, has an isotropic background and spectrum of perturbations, and  a (potentially) maximally anisotropic three-point function.

Finally, it would be interesting to run a dedicated numerical analysis of CMB data for our specific three-point function template, given its small overlaps with the more standard templates that have been considered so far. 
%It would also be interesting to understand whether there are large scale structure observables that are sensitive to our form of non-gaussianity.

We hope to address all these questions in the near future.

%%%%%%%%%%%%%%%%%%%%%%%%%%%%%%%%%%
%%%%%%%%%%%%%%%%%%%%%%%%%%%%%%%%%%
\section*{Acknowledgements}

We would like to thank Nima Arkani-Hamed, Sergei Dubovsky, Lam Hui, and Raman Sundrum for very useful discussions. We are especially thankful to Paolo Creminelli and Matias Zaldarriaga for their insightful comments.
The work of S.E.~is supported by the NSF through a Graduate Research Fellowship.
The work of A.N.~is supported by NASA ATP under contract NNX10AH14G and by the DOE under contract DE-FG02-11ER41743.
The work of J.W.~is supported by the DOE under contract DE-FG02-92-ER40699.

%%%%%%%%%%%%%%%%%%%%%%%%%%%%%%%%%%
%%%%%%%%%%%%%%%%%%%%%%%%%%%%%%%%%%

\appendix
\section*{Appendix}

%%%%%%%%%%%%%%%%%%%%%%%%%%%%%%%%%%%%%%%%%%%%%%%
%%%%%%%%%%%%%%%%%%%%%%%%%%%%%%%%%%%%%%%%%%%%%%%%
\section{Time dependence of background quantities}\label{time dep}
In order to solve the classical equations of motion for scalar and tensor perturbations, we need know the explicit time dependence of quantities such $a(\tau)$, $H(\tau)$, $\ep(\tau)$, \dots; the goal of this section is obtaining this time dependence. For the computations we are interested in, it suffices to derive these temporal functions up to the first order in slow roll. 
To make the notation lighter, we will mostly  drop the $\tau$ argument: $a(\tau) \to a $, etc. Primes will denote derivatives with respect to $\tau$.

%Let's begin by defining a particular time $\tau_0$ (or $t_0$) such that $a(\tau_0)=1$. It follows then from (\ref{rho $and p}), (\ref{Friedman}) that the Hubble parameter at such time $H_0$ is given by some intrinsic parameter in our %Lagrangian:
%\be
%H_0=-\frac{F(3,\frac{1}{3},\frac{1}{9})}{3\mpl^2}
%\ee
%Aside from $\mpl$, $H_0$ is the only dimsionful input of our model. Eventually any dimensionful quantities, $\tau_0$ for instance, should be expressed in terms of powers of $H_0$ (and $\mpl$ if necessary).  

Recall the definition of the first slow roll parameter $\ep$, (\ref{slow roll}), and rewrite it as 
\be
\ep =-\frac{H' }{a H^2}=\frac{d}{d \tau}\left(\frac{1}{a H }\right)+1 \;.
\ee
Integrating the above equation once and choosing some suitable additive constant
\footnote{The integration constant is chosen by demanding $a(\tau)\gg a(\tau_c)$, for $\tau/\tau_c \to 0$. }, one has 
\be
\frac{1}{a H}=-(1-\ep_c) \, \tau+\mathcal{O}(\ep^2)\label{H0 tau0}
\ee
where the subscript ``$c$'' denotes evaluation at some reference conformal time $\tau_c$, which we find most convenient to choose to be the (conformal) time when the {\it longest} mode of observational relevance today exists the horizon, i.e. $\vert c_{L,c}k_{\rm min}\tau_c\vert \simeq \vert c_{L,c} \tau_c H_{\rm today}\vert=1
$. 

The reason $\ep(\tau)$ is being treated as a constant in integration is that the higher order terms in the Taylor series of $\ep(\tau)=\ep(\tau_c)+\ep'(\tau_c)(\tau-\tau_c)+\dots$ are suppressed by powers of slow roll, for instance $\ep'(\tau_c) \tau_c \sim O(\ep^2)$.  Of course this also depends on the choice of reference time; we don't want the perturbative expansion in slow roll to be contaminated by large values of $\tau/\tau_c-1$. Since $\vert \tau(t) \vert$ is a decreasing function during inflation, and $+\infty>-\tau>0$, we ought to choose early times (like $\tau_c$) as the reference.

Using the definition of the Hubble parameter, we can extract the time dependence of the scale factor $a(\tau)$ from the above equation (\ref{H0 tau0}):
\be
a H=\frac{a' }{a }=-\frac{1+\ep_c}{\tau}\quad\Longrightarrow \quad a(\tau)= a_c \left(\frac{\tau}{\tau_c}\right)^{-1-\ep_c}+\mathcal{O}\;.(\ep^2)\label{a of tau}
\ee
Furthermore, we obtain
\be
H (\tau) =\frac{a'}{a^2}=-\frac{1+\ep_c}{a_c \tau_c}\left(\frac{\tau}{\tau_c}\right)^{\ep_c}+\mathcal{O}(\ep^2)\;.
\ee
Finally, the time dependence of $\ep$, $c_L$ and $c_T$ can be revealed by invoking the definitions of other slow roll parameters. For instance,
\be
\frac{\ep'}{\ep}=aH\eta=-\frac{\eta}{\tau}+\mathcal{O}(\ep^2) \quad \Longrightarrow \quad  \ep(\tau)=\ep_c \left(\frac{\tau}{\tau_c}\right)^{-\eta_c}+\mathcal{O}(\ep^3) \;.
\ee
Similarly we obtain
\be
c_L(\tau)=c_{L,c} \left(\frac{\tau}{\tau_c}\right)^{-s_c}+\mathcal{O}(\ep^2),\qquad c_L(\tau)=c_{T,c} \left(\frac{\tau}{\tau_c}\right)^{-u_c}+\mathcal{O}(\ep^2)\label{c of tau} \;.
\ee
Notice that, because of the all-order relation between $c_T^2$ and $c_L^2$ of footnote 4, $c_{T,c}$ and $u_c$ are not independent parameters---they can be expressed in terms of $c_{L,c}$ and of the slow-roll parameters.
The equations (\ref{a of tau})--(\ref{c of tau}) are frequently used in solving the classical equations of motion.

%%%%%%%%%%%%%%%%%%%%%%%%%%%%%%%%%%%%%%%%%%%%%%%%%
%%%%%%%%%%%%%%%%%%%%%%%%%%%%%%%%%%%%%%%%%%%%%%%%

\section{Unitary gauge vs.~spatially flat slicing gauge}
\label{gauge app}
In performing calculations throughout this paper two gauge choices are particularly useful:
\begin{itemize}
\item Spatially Flat Slicing Gauge (SFSG) is defined by setting to zero the scalar and vector perturbations in $g_{ij}$, i.e.~by imposing
\be 
g_{ij}=a(t)^2 \exp{\gamma}_{ij} \;,
\ee
where $\gamma_{ij}$ denotes the transverse traceless tensor mode, satisfying 
\be
\gamma_{ii}=0 \; , \qquad \pd_i \gamma_{ij}=0 \; . 
\ee
Then the fluctuations in our $\phi^I$ scalars are unconstrained:
\be
\phi^I = x^I + \pi^I \; .
\ee
The three $\pi^I(x)$ fields can be split into a transverse vector and longitudinal scalar as in (\ref{decomposing pi}), according to their transformation properties under the {\em residual} rotation group. This gauge choice is of particular convenience for computations of the two- and three-point functions because in the demixing (with gravity) limit the $\pi$ Lagrangian will contain all the scalar (or longitudinal) and vector (or transverse) degrees of freedom. 

\item Unitary Gauge (UG) is defined by setting to zero the fluctuations in the $\phi^I$ fields and in the ``clock'' field:  
\be
\phi^I =x^I, \qquad  \det(B^{IJ})=a(t)^{-6} \;. \label{UG}
\ee
Then the spatial metric is unconstrained. And can be parameterized in general as
\be
g_{ij}=a(t)^2\exp(A \, \delta_{ij}+\pd_i\pd_j \chi+\pd_i C_j+\pd_j C_i +D_{ij})  \; ,
\ee
where $\di_i C_i=0$ and $D_{ij}$ is transverse-traceless.
\end{itemize}
 From the above form of the metric it seems that in UG there are too many degrees of freedom; there is an extra scalar in addition to the scalar, transverse vector, and transverse traceless tensor that we expect. However, when the metric is expressed in terms of the ADM parameters defined by (\ref{ADM decomposition}), the second condition in (\ref{UG}) can be rewritten as 
\be
3A+\nabla^2 \chi=\log(1-N^i N_i/N^2) \;. \label{UG2}
\ee

As $N^i$ and $N$ can be expressed in terms of $A$, $\chi$, $C_i$, and $D_{ij}$ by solving the constraint equations given by (\ref{N constraint}), (\ref{N^i constraint}) we can see that (\ref{UG2}) implies that the two scalar functions $A(x)$ and $\chi(x)$ are not independent in UG. Hence the dynamical d.o.f.~in question can be chosen to be $A(x)$ (the scalar mode), $C_i(x)$ (the transverse vector mode), and $D_{ij}$ (the transverse traceless tensor mode). The number of which matches our physical intuition and properly agrees with SFSG. UG is particularly useful in following our degrees of freedom through the reheating surface.

As we find it convenient to calculate correlation functions in SFSG, and yet utilize UG to most easily describe the surface of sudden reheating, we need to develop the transformation rules to go from one gauge to the other. Let's denote by $\{x^\mu\}$ the coordinate system in SFSG and by $\{\bar{x}^\mu\}$ that in UG. A gauge transformation relating SFSG to UG is given by $\bar{x}^\mu=x^\mu+\xi^\mu(x)$,
where
\begin{align}
\xi^0(x)=-\frac{1}{3H}\pd_i\pi^i(x)+\mathcal{O}(\pi^2), \quad \xi^I(x)=\pi^I(x)+\mathcal{O}(\pi^2)
\end{align}
and the scalar perturbations are related by 
\be
A=\frac{2}{3}\pd_i\pi^i+\mathcal{O}(\pi^2) \;.
\ee
%%%%%%%%%%%%%%%%%%%%%%%%%%%%%%%%%%%%%%%%%%%%%%%%%%
%%%%%%%%%%%%%%%%%%%%%%%%%%%%%%%%%%%%%%%%%%%%%%%%%%%%%

\section{Vector perturbations}\label{vectors}
%In our solid inflation model,  we begin with one longitudinal scalar mode and one transverse vector mode (for instance, in SFSG, they are characterized by $\pi_L$ and $\pi_T^i$).  Since the transverse vector perturbations play secondary role in the post-inflationary cosmology, we don't bother to compute their $2-$pt correlation function or their tilt.  Nevertheless, it is relevant to ask how large vector perturbations are at the end of inflation, after all,  large vector perturbations will leave imprints on the CMB map and hence potentially can be detected. 

In this section we will derive the spectrum for  the vector modes $\pi_T^i$ (in SFSG), in the same manner as for scalar perturbations and tensor perturbations.  Let's begin by writing the vector modes as 
\be
\pi_T^i(\vec k,t)=\sum_{\lambda=\pm}\epsilon^{i}_{\lambda}(\vec{k}) \, \pi_{T,\lambda} (\vec k,t) \;,
\ee
where the polarization vectors satisfy the transverse condition $k_i\ep^i(\vec k)=0$ and they form an orthonormal and complete set $\ep^i_{\lambda}(\vec k)\ep^i_{\lambda'}(\vec k)^*=\delta_{\lambda \lambda'}$. As before, writing
\be
\pi_{T,\lambda}(\vec k, t)=\pi^{cl}_T(\vec k, t) \; d_\lambda(\vec k)+\pi^{cl}_T(\vec k,t)^{*} \; d_{\lambda}^{\dag}(-\vec k)
\ee
the creation and annihilation operators obey the usual commutation relation $[d_\lambda(\vec k), d_{\lambda'}(\vec{k}')^\dag]=(2\pi)^3 \delta_{\lambda \lambda'}\delta^{3}(\vec{k}-\vec{k}')$\; and the classical equations of motions for $\pi_T^{cl}$ follow from varying the quadratic $\pi_T$ action (\ref{transverse quadratic action}), which is given by
\begin{align}
&\dot{N}_T^{cl}+3 H N_T^{cl}+4H^2\ep\, c_T^2\pi_T^{cl}=0 \label{eomvector1}\\
&N_T^{cl}=\frac{\dot{\pi}_T^{cl}}{1+k^2/4a^2H^2\ep} \label{eomvector2}
\end{align}
where the second one follows from (\ref{NT}) (after $N_T^i$ is expressed in terms creation and annihilation operators in the same manner as $\pi_T^i$). Eliminating $\pi_T^{cl}$, we reach a second order differential equation for $N_T^{cl}$; using conformal time and keeping terms only up to the first order in slow roll, it reads
\be
\frac{d^2}{d \tau^2}N_T^{cl}-\frac{2+4\ep_c-\eta_c-2u_c}{\tau}\frac{d}{d\tau}N_T^{cl}+\left[\frac{3\ep_c-3\eta_c-6u_c+4c_{T,c}^2\ep_c}{\tau^2}+k^2 c_T(\tau)^2\right]N_T^{cl}=0
\ee
where $u\equiv \dot{c}_T/H c_T=\frac{3 c_{L,c}^2}{4 c_{T,c}^2}s+O(\ep^2)$. The general solution for the equation above takes the form
\begin{align}
N_T^{cl}&=(-\tau)^\beta \left[\mathcal{E} H^{(1)}_{\nu_V} \left((1+u_c)c_{T,c}\, k \vert \tau_c\vert \left(\frac{\tau}{\tau_c}\right)^{1-u_c} \right) +\mathcal{F} H^{(2)}_{\nu_V} \left((1+u_c)c_{T,c}\, k \vert \tau_c\vert \left(\frac{\tau}{\tau_c}\right)^{1-u_c} \right) \right]\nonumber\\
&= \left(-\tau \right)^{-\beta} \left[\mathcal{E} H^{(1)}_{\nu_V} \left(- c_T(\tau) k \tau (1+u_c) \right) +\mathcal{F} H^{(2)}_{\nu_V} \left(- c_T(\tau) k \tau (1+u_c) \right) \right] \label{NT classical}
\end{align}
where $\beta=-\frac{3}{2}-2\ep_c+\frac12{\eta_c}+u_c$ and $\nu_V=\frac{3}{2}+\frac52{u_c}-c_{L,c}^2\ep_c+\frac12{\eta_c}$.  The constants $\mathcal{E}$ and $\mathcal{F}$ are determined via the initial condition, which is specified by 
\begin{align}
\lim_{\tau\to -\infty}N_T^{cl}(\vec k, \tau)=\lim_{\tau\to -\infty}\frac{4 H^2 a \ep}{k^2}\frac{d}{d \tau}\pi_T^{cl}=-i \frac{\sqrt{4\ep c_T }H}{\mpl a} \frac1 { k^{3/2}}e^{-i c_T(\tau) \, k\tau(1+u_c)}
\end{align}
where in the second equality we have used the fact that the mode function of the canonically normalized transverse vector field $\pi^{\rm can.}_{T,cl}$ should match that in the flat space vacuum at early times, that is
\begin{align}
\pi_T^{cl}=  \left[\frac{ 2 a^2 \mpl^2 k^2 }{4\left(1+\frac{k^2}{4a^2 H^2 \epsilon}\right)}\right]^{-1/2}\pi_{T,cl}^{\rm can.} \stackrel{k \tau \to -\infty}{\longrightarrow} \frac{1}{\sqrt{2\ep} \, \mpl H a ^2} \frac{e^{-i c_T(\tau)k \tau (1+u_c)}}{\sqrt{2c_T k}} \; .
\end{align}
Therefore, $\mathcal{F}=0$ and 
\be
\mathcal{E}=-i \sqrt{2\pi \ep_c} \, \frac{c_{T,c}H_c^2}{k \mpl}\left(1+\frac{u_c}{2}-\ep_c\right)e^{\frac{i\pi}{2}\left(\nu_V+1/2\right)}\left(-\tau_c\right)^{-2\ep_c+\eta_c/2+u_c}+O(\ep^{5/2}) \; .
\ee
Then, by (\ref{eomvector1}), we can derive the expression for $\pi_T^{cl}$; in particular, its asymptotic behavior at very late time when the mode exits the horizon is given by 
\be
\lim_{k\tau\to 0^-}\pi^{cl}_T(\vec k,\tau)=\left(\frac{\tau}{\tau_c}\right)^{\frac{4}{3}c_{T,c}^2\ep_c}(-c_{T,c} \, k \, \tau_c)^{c_{L,c}^2\ep_c-\frac{\eta_c}{2}-\frac{5u_c}{2}}\left(\frac{-3H_c}{\sqrt{4\ep_c}\mpl c_{T,c}^{5/2}k^{5/2}}+O(\ep^{1/2})\right) \; ,
\ee
the time dependence of which is the {\it same} as the tensor (\ref{asym tensor}) and scalar (\ref{asym scalar2}) perturbations.  
The vector-to-scalar ratio simply
\be
\lim_{k\tau\to 0^-} \frac{ \big| \pi^{cl}_T(\vec k,\tau) \big|^2 }{\big|\pi_L^{cl}(\vec k, \tau) \big|^2} =\bigg(\frac{c_{L,c}^2}{c_{T,c}^2}\bigg)^{5/2}+\mathcal{O}(\ep) \; .
\ee
Given the relation \eqref{cT}, this is always smaller than $(1/3)^{5/2} \simeq 6\%$, which is reached only in the ultra-relativistic case with 
$c_L^2 \simeq 1/3$ and $c_T^2 \simeq1$, or equivalently when $\vert F_Y+F_Z\vert \sim \mathcal{O}(\ep^2)\vert F\vert$.

%%%%%%%%%%%%%%%%%%%%%%%%%%%%%%%%%%%%%%%%%%%%%%%%%%%%%%%
%%%%%%%%%%%%%%%%%%%%%%%%%%%%%%%%%%%%%%%%%%%%%%%%%%%%%

\section{Full trilinear action in SFSG}\label{trilinear action}
Expanding the Lagrangian to third order in fluctuations about the FRW background in SFSG we have after a straightforward but lengthy computation
\footnote{Since we are after the three-point function for scalar perturbations, we ignore the interaction between the tensor mode $\gamma$ and the $\pi$ fields.}
\begin{align}
\mathcal{L}^{(3)}&=a(t)^3\bigg\{ 3\mpl^2 H^2\stackrel{(1)}{\delta N^3}
+2\mpl^2 H \dNi \stackrel{(1)}{\delta N^2}+2\mpl^2\dot{H}a^2\stackrel{(1)}{N^j}\pd_j\pi^i(\dot{\pi}^i-\stackrel{(1)}{N^i})\nonumber\\
&\quad-\mpl^2 \dN \Big(\sfrac{1}{4}\stackrel{(1)}{\pd_i N^j}\stackrel{(1)}{\pd_j N^i}
+\sfrac{1}{4}\stackrel{(1)}{\pd_j N^i}\stackrel{(1)}{\pd_j N^i}
-\sfrac12(\dNi)^2\Big)\nonumber\\
&-\dN \mpl^2\dot{H}\Big(-a^2(\dot{\pi}^i-\stackrel{(1)}{N^i})^2-c_T^2[\Pi^T\Pi]+(1-c_T^2)[\Pi^2]-(1+c_L^2-2c_T^2)[\Pi]^2\Big)\nonumber\\
&+[\Pi]^3\left(\sfrac{4}{3}F_{XXX}a^{-6}-\sfrac{8}{27}(F_{XZ}+F_{XY})a^{-2}+\sfrac{64}{243}F_Z+\sfrac{16}{81}F_Y\right)\nonumber\\
&+[\Pi][\Pi^2]\left(\sfrac{4}{9}(F_{XZ}+F_{XY})a^{-2}-\sfrac{4}{9}F_Z-\sfrac{8}{27}F_Y\right)\nonumber\\
&+[\Pi][\Pi^T\Pi]\left(2F_{XX}a^{-4}+\sfrac{4}{9}(F_{XZ}+F_{XY})a^{-2}-\sfrac{16}{27}F_Z-\sfrac{4}{9}F_Y\right)\nonumber\\
&+\sfrac{2}{27}F_Z[\Pi^3]+\left(\sfrac{2}{3}F_Z+\sfrac{4}{9}F_Y\right)[\Pi^T\Pi\Pi]
-\sfrac{4}{9}(F_Y+F_Z)a^2 (\dot{\pi}^i-\stackrel{(1)}{N^i})\pd_i\pi^j(\dot{\pi}^j-\stackrel{(1)}{N^j})\nonumber\\
&+a^2[\Pi](\dot{\pi}^i-\stackrel{(1)}{N^i})^2\left(-2F_{XX}a^{-4}+\sfrac{4}{27}(F_Y+F_Z)\right)
\bigg\} \;,
\end{align}
where $\Pi$ denotes the $3\times3$ matrix $\pd_i \pi^j$ and $[\cdots]$ indicates the trace; for instance, $[\Pi^T\Pi\Pi]\equiv \pd_j\pi^i\pd_j\pi^k\pd_k\pi^i$.

Now, as discussed in Section \ref{Expanding in perturbations} one only needs to solve the constraint equations  $\delta S/\delta N=0$ and $\delta S/\delta N^i=0$ to linear order in perturbations. We therefore don't need to worry about terms in $N$, $N^i$ that are quadratic in the fluctuations contributing to the cubic Lagrangian. The solutions to these equations are given by (\ref{delta N}), (\ref{N_L}), and (\ref{N_T}). 

In particular, we are interested in two separate limits for computing the three-point function. The first is the de-mixing regime where $k\gg aH \epsilon^{1/2}$. In this limit, to lowest order in slow roll, we can effectively set $\delta N$ and $N_L$ to zero. Furthermore, note that all terms that are not of the form $\Pi^3$ (like the final $[\Pi] \dot{\pi}^2$ term) are explicitly suppressed by slow roll parameters. We are left with
\be
\label{L_3 demixed}
\mathcal{L}_{3}= a(t)^3 \mpl^2 H^2\frac{F_Y}{F}\Big\{\sfrac{7}{81}(\pd \pi)^3-\sfrac{1}{9}\pd\pi \pd_j\pi^k\pd_k\pi^j-\sfrac{4}{9}\pd\pi \pd_j\pi^k\pd_j\pi^k +\sfrac{2}{3}\pd_j\pi^i \pd_j\pi^k\pd_k\pi^i\Big\}  \; ,
\ee
where we neglected boundary terms.
We have freely used the Friedmann equations (\ref{Friedman}), various definitions of slow roll parameters, and the the total derivative 
\be
\det (\di \pi)=\sfrac{1}{6} \epsilon_{ijk} \epsilon_{lmn} \di^i \pi^l \di^j \pi^m \di^k \pi^n=\sfrac{1}{6} \left([\di \pi]^3-3[\di \pi][\di \pi ^2]+2[\di \pi^3] \right) 
\ee
(which is a total derivative because of the $\ep$-tensor structure).

The second limit is in the strong mixing (with gravity) regime. This occurs when $k \ll aH \epsilon^{1/2}$. In this limit we can write to lowest order in slow roll
\bea
\delta N &=& \frac{k}{a}\frac{d}{d\tau} \left(\frac{\pi_L}{H}\right)\simeq k \epsilon \, \pi_L \;,\\
N_L &=&\dot{\pi}_L \;,\\
N_T^i &=& \dot{\pi}_T ^i\;,
\eea
where we have estimated the time dependence of $\pi$ via the explicit classical solution to the first order equation of motion given by (\ref{pidemix}). When we insert the above expressions into the full cubic Lagrangian we see that all the terms involving these auxiliary fields are going to vanish, as the reccurring combination $(\dot{\pi}^i-N^i)$ vanishes to lowest order in slow roll and $\delta N$ is explicitly of order $\epsilon (\di \pi)$. So, surprisingly, we see that to lowest order in slow roll we recover the same expression (\ref{L_3 demixed}) for the cubic Lagrangian in the {\em strong} mixing limit. This is a convenient fact, as it allows us to effectively use the same expression for the cubic interactions during the whole inflationary phase in our calculation of the three-point function in Section \ref{3ptfunction}.

%%%%%%%%%%%%%%%%%%%%%%%%%%%%%%%%%%
%%%%%%%%%%%%%%%%%%%%%%%%%%%%%%%%%%

\section{Three point function integral for general triangles}\label{for general triangles}

For momenta configurations that are not roughly equilateral, we need to be a little more careful in computing the integral \eqref{Def of Integral}, as there are many different regions in the integral, each of which will necessitate the use of different expressions of $\pi_L(\tau, k)$ given by (\ref{pidemix}) and (\ref{pimix}).

Assume for the general case that $k_1> k_2 >k_3$ and define three separate times $\tau_1$, $\tau_2$, and $\tau_3$ by
\be
-c_L k_1 \tau_1=\epsilon \; , \qquad -c_L k_2 \tau_2=\epsilon \; , \qquad  \quad -c_L k_3 \tau_3=\epsilon \:, 
\ee
and for the observation time $\tau$ require $|\tau|<|\tau_1|$---that is, all modes are very long at the time of observation.

Splitting the object in question into regions defined by the times given above we can write schematically:
\bea
\nonumber
I(\tau; -\infty)&=&\prod_{i=1}^{3} \pi_L(\tau,  k _i ) \cdot \left (  \int_{-\infty}^{\tau_3}\; d\tau ' \quad +\int_{\tau_3}^{\tau_2}\; d\tau ' \quad +\int_{\tau_2}^{\tau_1}\; d\tau ' \quad +\int_{\tau_1}^{\tau}\; d\tau ' \right )\\
\label{split integral}
&& \times \left(\frac{-i}{H_0^2 \tau'^4}\right)\prod_{j=1}^3 \pi_L^*(\tau',  k _j ) + \text{c.c.}
\eea
We can do each integral separately, and will find that provided that the the triangle does not become ``too squeezed'' (which will become precise in a moment) only the first integral contributes to leading order in slow roll. For each integral there is, of course, going to be an overall $H_0^{-2}\prod_{i=1}^3 |\mathcal{B}_{k_i}|^2(-c_L k_i \tau)^{c_L^2\epsilon+\epsilon}$ as $|\tau|<|\tau_1|$. 

\subsection{Integrating from $-\infty$ to $\tau_3$:}

Ignoring for a moment the overall factor, we need to compute
\be
\int_{-\infty}^{\tau_3} d \tau' \; \left(\frac{-i}{\tau'^{4}}\right) \prod_{i=1}^3 \left(1-i c_L k_i \tau' -\frac{1}{3} c_L^2 k_i^2 \tau'^2 \right) e^{i c_L k_i \tau'}+\text{c.c} \equiv I_3 \; .
\ee
In order to ensure convergence of the integral and project onto the right vacuum the integral is computed over a slightly tilted contour, that is $\tau' \rightarrow (1-i \varepsilon )(\tau '+\tau_3)$, with $\varepsilon>0$, and the limits of integration are from $-\infty$ to $0$. Performing the integral and expanding in $\tau_3$ we have
\bea 
\nonumber
I_3 &=&-\frac{c_L^3 k_1 k_2 k_3 U(k_1,k_2, k_3)}{27}+\frac{c_L^5}{45}\left(k_1^5+k_2^5+k_3^5 \right)\cdot \tau_3^2\\
&&+\frac{c_L^7}{3780}\left(-3(k_1^7+k_2^7+k_3^7)+7(k_1^5 k_2^2+ \text{5 perms}) \right) \cdot \tau_3^4 +\mathcal{O}(\tau_3^6) \;,
\eea
where $U(k_1,k_2,k_3)$ is the scale invariant function given by (\ref{U}). Now, as $\tau_3=-\epsilon/c_Lk_3$ we can see that provided $k_1/k_3<(\sqrt{\epsilon})^{-1}$ only the zeroth order term in $\tau_3$ contributes to leading order in slow roll. That is, for a not ``too squeezed'' triangle the first integral in (\ref{split integral}) is given by
\be
-\frac{c_L^3 k_1 k_2 k_3 U(k_1,k_2, k_3)}{27 H_0^2 } \prod_{i=1}^3 |\mathcal{B}_{k_i}|^2 \times \left(1+\mathcal{O}(\epsilon) \right) \; , \quad \text{provided } \frac{k_1}{k_3}<\frac{1}{\sqrt{\epsilon}}\; .
\ee

\subsection{Integrating from $\tau_3$ to $\tau_2$:}

Again, dropping the overall factor, first note that
\bea
\nonumber
&&\int_{\tau_3}^{\tau_2} d \tau' \; \left(\frac{-i}{\tau'^{4}}\right)\left((-c_L k_3 \tau')^{\epsilon c_L^2+\epsilon} \right)\prod_{i=1}^2 \left(1-i c_L k_i \tau' -\frac{1}{3} c_L^2 k_i^2 \tau'^2 \right) e^{i c_L k_i \tau'}+\text{c.c} \\
\nonumber
&<&\left((-c_L k_3 \tau_3)^{\epsilon c_L^2+\epsilon} \right) \left[ \int_{\tau_3}^{\tau_2} d \tau' \; \left(\frac{-i}{\tau'^{4}}\right)\prod_{i=1}^2 \left(1-i c_L k_i \tau' -\frac{1}{3} c_L^2 k_i^2 \tau'^2 \right) e^{i c_L k_i \tau'}+\text{c.c} \right]\\
&\simeq&\int_{\tau_3}^{\tau_2} d \tau' \; \left(\frac{-i}{\tau'^{4}}\right)\prod_{i=1}^2 \left(1-i c_L k_i \tau' -\frac{1}{3} c_L^2 k_i^2 \tau'^2 \right) e^{i c_L k_i \tau'}+\text{c.c} \equiv I_2 \; .
\eea
This final form of the integral is straightforward to compute and yields
\be
I_2=-\frac{c_L^3(k_1^5+k_2^5)(k_2-k_3)(k_2+k_3)}{45 k_2^2 k_3^2} \cdot \epsilon^2 +\mathcal{O}(\epsilon^4)
\ee
which, provided that $k_1/k_3<(\sqrt{\epsilon})^{-1}$ is first order in slow roll. That is, for a not ``too squeezed'' triangle the second integral in (\ref{split integral}) is given by
\be
\frac{c_L^3 k^3}{ H_0^2 } \prod_{i=1}^3 |\mathcal{B}_{k_i}|^2 \times \mathcal{O}(\epsilon) \; , \quad \text{provided } \frac{k_1}{k_3}<\frac{1}{\sqrt{\epsilon}}\; .
\ee

\subsection{Integrating from $\tau_2$ to $\tau_1$:}

The exact same logic follows for this piece. The meat of the integral,
\be
\int_{\tau_2}^{\tau_1} d \tau' \; \left(\frac{-i}{\tau'^{4}}\right) \left(1-i c_L k_1 \tau' -\frac{1}{3} c_L^2 k_1^2 \tau'^2 \right) e^{i c_L k_1 \tau'}+\text{c.c}\equiv I_1\; ,
\ee
yields
\be
I_1=\frac{c_L^3}{45}k_1^3\left(1-\frac{k_1^2}{k_2^2} \right) \cdot \epsilon^2 +\mathcal{O}(\epsilon^4)\; .
\ee
Which, provided that $k_1/k_2<(\sqrt{\epsilon})^{-1}$ is first order in slow roll. And so, once again, for a not ``too squeezed'' triangle the third integral in (\ref{split integral}) is given by
\be
\frac{c_L^3 k^3}{ H_0^2 } \prod_{i=1}^3 |\mathcal{B}_{k_i}|^2 \times \mathcal{O}(\epsilon) \; , \quad \text{provided } \frac{k_1}{k_2}<\frac{1}{\sqrt{\epsilon}}\; .
\ee

\subsection{Integrating from $\tau_1$ to $\tau$:}

As done in the main body of the text, we once can immediately see that this integral is going to be of the form:
\be
\prod_{i=1}^3 \left|\mathcal{B}_{k_i}\right|^2 (-c_L k_i \tau)^{c_L^2\ep+\epsilon}\left[ \int_{\tau_1}^\tau d \tau'\; (-i)g(\tau') +\text{c.c.}\right] 
\ee
where $g(\tau)$ is pure real. And thus, the final integral in (\ref{split integral}) simply vanishes.

\subsection{Summary:}
And so, for not ``too squeezed'' triangles, $I(\tau; -\infty)$ is completely dominated by the portion of the integral where all the modes are in the de-mixed regime, that is:
\be
I(\tau; -\infty)=
-\frac{c_L^3 k_1 k_2 k_3 U(k_1,k_2, k_3)}{27 H_0^2 } \prod_{i=1}^3 |\mathcal{B}_{k_i}|^2 \times \left(1+\mathcal{O}(\epsilon) \right) \; , \quad \text{provided } \frac{k_i}{k_j}>\sqrt{\epsilon} \quad \; .
\ee

%%%%%%%%%%%%%%%%%%%%%%%%%%%%%%%%%%%%%%%%%%%
%%%%%%%%%%%%%%%%%%%%%%%%%%%%%%%%%%%%%%%%%%%


\begin{thebibliography}{99}
\small


%\cite{Creminelli:2006xe}
\bibitem{CLNS}
P.~Creminelli, M.~A.~Luty, A.~Nicolis and L.~Senatore,
``Starting the Universe: Stable Violation of the Null Energy Condition and Non-Standard Cosmologies,''
JHEP {\bf 0612} (2006) 080
[arXiv:hep-th/0606090].
%%CITATION = JHEPA,0612,080;%%


%\cite{Cheung:2007st}
\bibitem{CCFKS}
C.~Cheung, P.~Creminelli, A.~L.~Fitzpatrick, J.~Kaplan and L.~Senatore,
``The Effective Field Theory of Inflation,''
JHEP {\bf 0803} (2008) 014
[arXiv:0709.0293 [hep-th]].
%%CITATION = JHEPA,0803,014;%%



%\cite{Gruzinov:2004ty}
\bibitem{Gruzinov}
A.~Gruzinov,
``Elastic Inflation,''
Phys.\ Rev.\ D {\bf 70} (2004) 063518
[arXiv:astro-ph/0404548].
%%CITATION = PHRVA,D70,063518;%%

\bibitem{BCZ} 
  D.~Babich, P.~Creminelli and M.~Zaldarriaga,
  ``The Shape of non-Gaussianities,''
  JCAP {\bf 0408}, 009 (2004)
  [astro-ph/0405356].
  %%CITATION = ASTRO-PH/0405356;%%

\bibitem{BS} 
  M.~Bucher and D.~N.~Spergel,
  ``Is the dark matter a solid?,''
  Phys.\ Rev.\ D {\bf 60}, 043505 (1999)
  [astro-ph/9812022].
  %%CITATION = ASTRO-PH/9812022;%%

\bibitem{GHKRT} 
  P.~W.~Graham, B.~Horn, S.~Kachru, S.~Rajendran and G.~Torroba,
  ``A Simple Harmonic Universe,''
  arXiv:1109.0282 [hep-th].
  %%CITATION = ARXIV:1109.0282;%%

\bibitem{Leutwyler} 
  H.~Leutwyler,
  ``Phonons as goldstone bosons,''
  Helv.\ Phys.\ Acta {\bf 70}, 275 (1997)
  [hep-ph/9609466].
  %%CITATION = HEP-PH/9609466;%%

\bibitem{DGNR} 
  S.~Dubovsky, T.~Gregoire, A.~Nicolis and R.~Rattazzi,
  ``Null energy condition and superluminal propagation,''
  JHEP {\bf 0603}, 025 (2006)
  [hep-th/0512260].
  %%CITATION = HEP-TH/0512260;%%

\bibitem{ENRW} 
  S.~Endlich, A.~Nicolis, R.~Rattazzi and J.~Wang,
  ``The Quantum mechanics of perfect fluids,''
  JHEP {\bf 1104}, 102 (2011)
  [arXiv:1011.6396 [hep-th]].
  %%CITATION = ARXIV:1011.6396;%%

\bibitem{Son} 
  D.~T.~Son,
  ``Effective Lagrangian and topological interactions in supersolids,''
  Phys.\ Rev.\ Lett.\  {\bf 94}, 175301 (2005)
  [cond-mat/0501658].
  %%CITATION = COND-MAT/0501658;%%

\bibitem{RZ} 
  R.~Rattazzi and A.~Zaffaroni,
  ``Comments on the holographic picture of the Randall-Sundrum model,''
  JHEP {\bf 0104}, 021 (2001)
  [hep-th/0012248].
  %%CITATION = HEP-TH/0012248;%%

\bibitem{Polchinski} 
  J.~Polchinski,
  ``Scale And Conformal Invariance In Quantum Field Theory,''
  Nucl.\ Phys.\ B {\bf 303}, 226 (1988).
  %%CITATION = NUPHA,B303,226;%%

\bibitem{LPR} 
  M.~A.~Luty, J.~Polchinski and R.~Rattazzi,
  ``The $a$-theorem and the Asymptotics of 4D Quantum Field Theory,''
  arXiv:1204.5221 [hep-th].
  %%CITATION = ARXIV:1204.5221;%%

\bibitem{AADNR}
  A.~Adams, N.~Arkani-Hamed, S.~Dubovsky, A.~Nicolis and R.~Rattazzi,
  ``Causality, analyticity and an IR obstruction to UV completion,''
  JHEP {\bf 0610}, 014 (2006)
  [hep-th/0602178].
  %%CITATION = HEP-TH/0602178;%%

\bibitem{KW} 
  L.~-M.~Wang and M.~Kamionkowski,
  ``The Cosmic microwave background bispectrum and inflation,''
  Phys.\ Rev.\ D {\bf 61}, 063504 (2000)
  [astro-ph/9907431].
  %%CITATION = ASTRO-PH/9907431;%%

\bibitem{CNSTZ} 
  P.~Creminelli, A.~Nicolis, L.~Senatore, M.~Tegmark and M.~Zaldarriaga,
  ``Limits on non-gaussianities from wmap data,''
  JCAP {\bf 0605}, 004 (2006)
  [astro-ph/0509029].
  %%CITATION = ASTRO-PH/0509029;%%



\bibitem{Weinberg_Adiabatic_Modes}
S.~Weinberg,
``Adiabatic modes in cosmology,''
 Phys.Rev. D67 (2003) 123504
[astro-ph/0302326] 




\bibitem{Maldacena}
J.~ Maldacena,
``Non-Gaussian features of primordial fluctuations in single field inflationary models,''
JHEP 0305 (2003) 013
[arXiv:astro-ph/0210603v5]


\bibitem{Weinberg}
S.~Weinberg, 
``Cosmology," 

\bibitem{BST} 
  J.~M.~Bardeen, P.~J.~Steinhardt and M.~S.~Turner,
  ``Spontaneous Creation of Almost Scale - Free Density Perturbations in an Inflationary Universe,''
  Phys.\ Rev.\ D {\bf 28}, 679 (1983).
  %%CITATION = PHRVA,D28,679;%%



\bibitem{SSZ} 
  L.~Senatore, K.~M.~Smith and M.~Zaldarriaga,
  ``Non-Gaussianities in Single Field Inflation and their Optimal Limits from the WMAP 5-year Data,''
  JCAP {\bf 1001}, 028 (2010)
  [arXiv:0905.3746 [astro-ph.CO]].
  %%CITATION = ARXIV:0905.3746;%%

\bibitem{DHNS} 
  S.~Dubovsky, L.~Hui, A.~Nicolis and D.~T.~Son,
  ``Effective field theory for hydrodynamics: thermodynamics, and the derivative expansion,''
  Phys.\ Rev.\ D {\bf 85}, 085029 (2012)
  [arXiv:1107.0731 [hep-th]].
  %%CITATION = ARXIV:1107.0731;%%

\bibitem{ACMZ} 
  N.~Arkani-Hamed, P.~Creminelli, S.~Mukohyama and M.~Zaldarriaga,
  ``Ghost inflation,''
  JCAP {\bf 0404}, 001 (2004)
  [hep-th/0312100].
  %%CITATION = HEP-TH/0312100;%%

\bibitem{Nicolis} 
  A.~Nicolis,
  ``Low-energy effective field theory for finite-temperature relativistic superfluids,''
  arXiv:1108.2513 [hep-th].
  %%CITATION = ARXIV:1108.2513;%%

\bibitem{Dubovsky} 
  S.~L.~Dubovsky,
  ``Phases of massive gravity,''
  JHEP {\bf 0410}, 076 (2004)
  [hep-th/0409124].
  %%CITATION = HEP-TH/0409124;%%

\bibitem{CZ} 
  P.~Creminelli and M.~Zaldarriaga,
  ``Single field consistency relation for the 3-point function,''
  JCAP {\bf 0410}, 006 (2004)
  [astro-ph/0407059].
  %%CITATION = ASTRO-PH/0407059;%%

\bibitem{WeinbergCC} 
  S.~Weinberg,
  ``The Cosmological Constant Problem,''
  Rev.\ Mod.\ Phys.\  {\bf 61}, 1 (1989).
  %%CITATION = RMPHA,61,1;%%

\bibitem{LMS}
D.~H.~Lyth, K.~A.~Malik, M.~Sasaki
``A general proof of the conservation of the curvature perturbation''
JCAP 0505:004, 2005
[arXiv:astro-ph/0411220v3]

\bibitem{LF}
D.~Langlois, F.~Vernizzi
``Conserved nonlinear quantities in cosmology''
Phys.\ Rev.\ D {\bf 72} (2005) 103501
[arXiv:astro-ph/0509078v3]

\bibitem{AL}
A.~F.~Andreev and I.~M.~Lifshitz,
Sov.\ Phys.\ JETP {\bf 29} (1969) 1107.

\end{thebibliography}
\end{document}